# Non-dipole effects in angular distributions of secondary electrons in fast particle-atom scattering

M. Ya. Amusia[1,2], L. V. Chernysheva[2], and E. Z. Liverts[1]

[1]Racah Institute of Physics, the Hebrew University, Jerusalem 91904, Israel
[2]Ioffe Physical-Technical Institute, St.-Petersburg 194021, Russia

**Abstract**

We demonstrate that the angular distribution of electrons knocked out from an atom by a fast charge particle is determined not only by dipole but also by quadrupole transitions, the contribution of which can be considerably enhanced as compared to the case of photoionization.

To obtain these matrix elements one has to study the angular distribution of electrons emitted by the atom in its collision with a fast charged particle. The distribution has to be measured relative to the momentum $q$ transferred from the projectile to the target atom.

The situation is similar, but not identical to the photoionization studies, where the matrix elements of continuous spectrum atomic quadrupole transitions can be determined by measuring the so-called non-dipole angular anisotropy parameters of photoelectrons. However, they are suppressed as compared to the dipole matrix elements by the parameter $\omega R/c \ll 1$, where $\omega$ is the photon energy, $R$ is the ionized shell radius and $c$ is the speed of light. This suppression is controlled in fast electron-atom collisions, where the respective expansion parameter $\omega R/v \ll 1$ can be much bigger than $\omega R/c \ll 1$ since the speed of the incoming electron $v$ is much smaller than $c$.

We present not only general formulas, but also concrete results of calculations for noble gas atoms He, Ar and Xe. We have investigated their outer and subvalent subshells, as well as strongly collectivized $4d^{10}$ subshell in Xe. It appeared that even for the case of very small transferred momentum $q$, i.e. in the so-called optical limit the deviation from photoionization case is prominent.

## I. Introduction

About ten - fifteen years ago a lot of attention has been given to investigation of the so-called non-dipole parameters of the photoelectrons angular distribution (see [1-3] and references therein). It was understood that this is in fact the only way to reveal the contribution of quadrupole continuous spectrum matrix elements of atomic electrons that in the absolute cross photoionization cross-section are unobservable in the shadow of much bigger dipole contribution. To study non-dipole parameters high intensity sources of continuous spectrum electromagnetic radiation were used [4-7].

By the order of magnitude the ratio quadrupole-to-dipole matrix elements in photoionization is characterized by the parameter $\omega R/c$, where $\omega$ is the photon energy, $R$ is the ionized shell radius and $c$ is the speed of light. For photon energies up to several keV that includes ionization potential of the inner 1s subshell even for medium atoms, it is $\omega R/c \ll 1$. In the absolute cross-sections dipole and quadrupole terms do not interfere, so that the ratio of quadrupole to dipole contributions in the absolute cross section is given by the second power of the parameter $\omega R/c \ll 1$ and some of these terms are canceling each other. As to the angular distribution, it includes the dipole – quadrupole interference terms in the first power of $\omega R/c \ll 1$ and therefore the relative role of quadrupole terms are bigger.

Quite long ago fast charged particle inelastic scattering process was considered as a "synchrotron for poor" [8]. This notion reflects the fact that fast charge particle inelastic scattering is similar to photoionization, since it is mainly determined by the dipole contribution. But contrary to the photoionization case the ratio "quadrupole-to-dipole" contributions can be much bigger, since instead of $\omega R/c \ll 1$ they are determined by the parameter $\omega R/v$, where $v$ is the speed of the projectile. Since $1 \ll v \ll c$, the quadrupole term in inelastic scattering is relatively much bigger[1]. The transferred in collision momentum $q$ is not bound to the transferred energy $\omega$ by a relation similar to $\omega = aq$, with $a$ being a constant. Therefore the collision experiment gives an extra degree of freedom to control the atomic reaction to the transferred energy and linear moment. This stimulates the current research, the aim of which is to derive formulas for the angular anisotropy parameters of electrons emitted off the atom in its inelastic scattering with a fast charged projectile, and to perform calculations of these parameters as functions of $\omega$ and $q$.

In this paper, we suggest to investigate the cross-section of inelastic scattering upon atom and to study the angular distribution of the emitted electrons relative to the momentum $q$ transferred to the atom from the projectile to the atom. As it is known, fast charged particle inelastic scattering cross section is proportional to the so-called generalized oscillator strength GOS density. Therefore, we will concentrate in this paper on the GOS density angular distribution as a function of the direction of the atomic electron relative to the vector $\vec{q}$.

In our calculations we will not limit ourselves to the one electron Hartree-Fock approximation, but include multi-electron correlations in the frame of the random phase approximation with exchange (RPAE) that was successfully applied to studies of photoionization and fast electron scattering [9, 10].

## 2. Generalized oscillator strength angular distribution

The cross-section of fast electron inelastic scattering upon an atom with ionization of an electron of *nl* subshell can be presented as [11]

$$\frac{d^2\sigma_{nl}}{d\omega do} = \frac{2\sqrt{(E-\omega)}}{\sqrt{E}\omega q^2}\frac{dF_{nl}(q,\omega)}{d\omega}. \tag{1}$$

Here $dF_{nl}(q,\omega)/d\omega$ is the differential in the ionized electron energy $\varepsilon = \omega - I_{nl}$ so-called generalized oscillator strength (GOS) density, $I_{nl}$ is the *nl* subshell ionization potential.

The differential in the emission angle of the ionized electron with linear momentum $\vec{k}$ the density of GOS of the ionized electron from a subshell with principal number *n* and angular momentum *l* in one-electron approximation is given by the following formula [12]:

$$\frac{df_{nl}(q,\omega)}{d\Omega} = \frac{1}{2l+1}\frac{2\omega}{q^2}\sum_m \left|\langle nlms|\exp(i\vec{q}\vec{r})|\varepsilon \vec{k}s\rangle\right|^2. \tag{2}$$

Here $\Omega$ is the solid angle of the emitted electron, *m* is the angular momentum projection, *s* is the electron spin, $\vec{q} = \vec{p} - \vec{p}'$ with $\vec{p}$ and $\vec{p}'$ being the linear moments of the fast incoming and outgoing electrons determined by the initial *E* and final *E'* energies as $p = \sqrt{2ME}$ and $p' = \sqrt{2ME'}$, while $M$ is the projectile mass. Note that $\omega = E - E'$ and $\varepsilon = \omega - I_{nl}$ is the outgoing electron energy, $I_{nl}$ is the $nl$ subshell ionization potential.

---
[1] Atomic system of units is used in this paper: electron charge *e*, its mass *m* and Plank constant $\hbar$ being equal to 1, $e = m = \hbar = 1$

The values of $\omega$ are limited by the relation $0 \leq \omega \leq pq/M = q\sqrt{2E/M}$, contrary to $\omega = cq$ for the case of photo-effect. In order to consider the projectile as fast, its speed must be much higher than the speed of electrons in the ionized subshell, i.e. $\sqrt{2E/M} \gg R^{-1}$. The transferred to the atom momentum $q$ is considered as small if $qR \leq 1$. For electron as a projectile the mass is given by $M = 1$.

According to the well-known expansion of $\exp(i\vec{q}\vec{r})$, one has:

$$\exp(i\vec{q}\vec{r}) = \sum_{L=0}^{\infty} H_L(\vec{q}\vec{r}), \qquad (3)$$

where

$$H_L(\vec{q}\vec{r}) = 4\pi i^L j_L(qr) \sum_{M=-L}^{L} (-1)^M Y_L^M(\theta_{\vec{q}}, \varphi_{\vec{q}}) Y_L^{-M}(\theta_{\vec{r}}, \varphi_{\vec{r}}). \qquad (4)$$

Here $\theta_{\vec{q}}, \varphi_{\vec{q}}$ are the azimuth and polar angles of $\vec{q}$ and $\theta_{\vec{r}}, \varphi_{\vec{r}}$ are the same for $\vec{r}$. This representation corresponds to the definition of spherical harmonic in *Mathematica* [13], and Spherical Harmonic Addition Theorem [14].

In Eq.(4), $j_L(qr) = \sqrt{\pi/2qr} J_{L+1/2}(qr)$ are the spherical Bessel functions and $J_{L+1/2}(qr)$ are the Bessel functions of the first kind.

We suggest measuring the angular distribution of the emitted electrons relative to $\vec{q}$. It means that the z-axis coincides with the direction of $\vec{q}$ and in (4) one has to put $\theta_{\vec{q}} = \varphi_{\vec{q}} = 0$. Taking into account the following relation $Y_L^M(0,0) = \sqrt{(2L+1)/4\pi}\delta_{M0}$, it is obtained

$$H_L(\vec{q}\vec{r}) = 2\sqrt{\pi(2L+1)} i^L j_L(qr) Y_L^0(\theta_{\vec{r}}, \varphi_{\vec{r}}). \qquad (5)$$

Let us present the one-electron atomic wave functions of the initial $\psi_{nlms}(\vec{r})$ and final $\psi_{\varepsilon \vec{k} s}(\vec{r})$ states in the following way (see e.g. [12]):

$$\begin{aligned}
\psi_{nlms}(\vec{r}) &= R_{nl}(r) Y_l^m(\theta_{\vec{r}}, \varphi_{\vec{r}}) \chi_s, \\
\psi_{\varepsilon \vec{k} s}(\vec{r}) &= \frac{1}{2k} \sum_{l'=0}^{\infty} (2l'+1) i^{l'} R_{kl'}(r) P_{l'}(\cos\theta_{\vec{k}\vec{r}}) e^{-i\delta_{l'}} \chi_s = \\
&= \frac{2\pi}{k} \sum_{l'=0}^{\infty} \sum_{m'=-l'}^{l'} i^{l'} e^{-i\delta_{l'}} R_{kl'}(r) (-1)^{m'} Y_{l'}^{m'}(\theta_{\vec{k}}, \varphi_{\vec{k}}) Y_{l'}^{-m'}(\theta_{\vec{r}}, \varphi_{\vec{r}}) \chi_s.
\end{aligned} \qquad (6)$$

Using these wave functions, the following expressions for the matrix elements of the operator (4) are obtained:

$$\begin{aligned}
\langle nlms | H_L(\vec{q}\vec{r}) | \varepsilon \vec{k} s \rangle &= \frac{2\pi(2L+1)\sqrt{(2l+1)} i^L (-1)^m}{k} \times \\
&\sum_{l'=|L-l|}^{L+l} (-i)^{l'} e^{i\delta_{l'}} \sqrt{2l'+1} \begin{pmatrix} L & l & l' \\ 0 & 0 & 0 \end{pmatrix} \begin{pmatrix} L & l & l' \\ 0 & m & -m \end{pmatrix} Y_{l'}^{m'}(\theta_{\vec{k}}, \varphi_{\vec{k}}) g_{nl,kl',L}(q),
\end{aligned} \qquad (7)$$

where

$$g_{nl,kl',L}(q) \equiv \int_0^\infty R_{nl}(r) j_L(qr) R_{kl'}(r) r^2 dr. \qquad (8)$$

Since we have in mind ionization of a particular *nl* subshell, for simplicity of notation and due to energy conservation in the fast electron inelastic scattering process leading to $k = \sqrt{2(\omega - I_{nl})}$, let us introduce the following abbreviations $g_{nl,kl',L}(q) \equiv g_{kl'L}(q)$. With its help we have for differential in the outgoing electron angle GOS density of *nl* subshell $df_{nl}(q,\omega)/d\Omega$ the following relation

$$\frac{df_{nl}(q,\omega)}{d\Omega} = \sum_{L'L''} \frac{df_{nl}^{(L'L'')}(q,\omega)}{d\Omega} = \frac{\omega\pi}{(\omega - I_{nl})q^2} \sum_{L'L''} (2L'+1)(2L''+1) i^{L'-L''} \times$$
$$\sum_{l'=|L'-l|}^{L'+l} \sum_{l''=|L''-l|}^{L''+l} g_{kl'L'}(q) g_{kl''L''}(q) i^{l''-l'} (2l'+1)(2l''+1) e^{i(\delta_{l'} - \delta_{l''})} \begin{pmatrix} L' & l & l' \\ 0 & 0 & 0 \end{pmatrix} \begin{pmatrix} L'' & l & L'' \\ 0 & 0 & 0 \end{pmatrix} \qquad (9)$$
$$\sum_{L=|l'-l''|}^{l'+l''} P_L(\cos\theta)(-1)^{L+l}(2L+1) \begin{pmatrix} l' & L & l'' \\ 0 & 0 & 0 \end{pmatrix} \begin{pmatrix} L & L' & L'' \\ 0 & 0 & 0 \end{pmatrix} \begin{Bmatrix} L & L' & L'' \\ l & l'' & l' \end{Bmatrix}.$$

Here $P_L(\cos\theta)$ are the $L^{th}$ order Legendre polynomials that depend upon the angle between the transferred in collision momentum $\bar{q}$ and the outgoing atomic electron momentum $\bar{k}$.

The partial value of GOS $f_{nl}(q,\omega)$ is obtained from (9) by integrating over $d\Omega$, leading to the following expression:

$$f_{nl}(q,\omega) = \sum_{L'} f_{nl}^{(L')}(q,\omega) = \frac{4\omega\pi^2}{(\omega - I_{nl})q^2} \sum_{L'} (2L'+1) \sum_{l'=|L'-l|}^{L'+l} [g_{kl'L'}(q)]^2 (2l'+1) \begin{pmatrix} L' & l & l' \\ 0 & 0 & 0 \end{pmatrix}^2. \qquad (10)$$

These formulas can be easily generalized in order to include inter-electron correlations in the frame of RPAE. This is achieved substituting $g_{kl'L'}(q)$ by $\bar{G}_{kl'L'}(q)$ and the scattering phases $\delta_{l'}$ by $\bar{\delta}_{l'} = \delta_{l'} + \Delta_{l'}$, where the expressions $G_{kl'L'}(q) \equiv \bar{G}_{kl'L'}(q) \exp(i\Delta_{l'})$ are solutions of the RPAE set of equations [15]:

$$\langle \varepsilon l' | G^L(\omega,q) | nl \rangle = \langle \varepsilon l' | j_L(qr) | nl \rangle +$$
$$+ \left( \sum_{\varepsilon''l'' \leq F, \varepsilon'''l''' > F} - \sum_{\varepsilon''l'' < F, \varepsilon'''l''' \leq F} \right) \frac{\langle \varepsilon'''l''' | G^L(\omega,q) | \varepsilon''l'' \rangle \langle \varepsilon''l'', \varepsilon l' | U | \varepsilon'''l''', nl \rangle_L}{\omega - \varepsilon_{\varepsilon'''l'''} + \varepsilon_{\varepsilon''l''} + i\eta(1 - 2n_{\varepsilon''l''})}. \qquad (11)$$

Here $\leq F (> F)$ denotes summation over occupied (vacant) atomic levels in the target atom. Summation over vacant levels include integration over continuous spectrum, $n_{\varepsilon l}$ is the Fermi step function that is equal to 1 for $nl \leq F$ and 0 for $nl > F$; the Coulomb interelectron interaction matrix element is defined as $\langle \varepsilon''l'', \varepsilon l' | U | \varepsilon'''l''', nl \rangle_L = \langle \varepsilon''l'', \varepsilon l' | r_<^L / r_>^{L+1} | \varepsilon'''l''', nl \rangle$ - $\langle \varepsilon''l'', \varepsilon l' | r_<^L / r_>^{L+1} | nl, \varepsilon'''l''' \rangle$. In the latter formula notation $r_<(r_>)$ comes from the well-known expansion of the Coulomb inter-electron interaction:

$$\frac{1}{|\vec{r}_1-\vec{r}_2|}=\sum_{L=0}^{\infty}(2L+1)\frac{r_<^L}{r_>^{L+1}}P_L(\cos\theta). \tag{12}$$

The necessary details about solving (11) one can find in [16].

## 3. Detailed expressions

To compare the results obtained with known formulas for the photoionization with lowest order non-dipole corrections taken into account, let us consider so small $q$ that it is enough to take into account terms with $L', L'' \leq 2$. In this case, GOS angular distribution (8) can be presented similar to the photoionization [1] as

$$\frac{df_{nl}(q,\omega)}{d\Omega}=\frac{f_{nl}(q,\omega)}{4\pi}\left\{1-\frac{\beta_{nl}^{(in)}(\omega,q)}{2}P_2(\cos\theta)+q\left[\gamma_{nl}^{(in)}(\omega,q)P_1(\cos\theta)+\eta_{nl}^{(in)}(\omega,q)P_3(\cos\theta)+\right.\right.$$
$$\left.\left.\varsigma_{nl}^{(in)}(\omega,q)P_4(\cos\theta)\right]\right\}. \tag{13}$$

The obvious difference is the $q$ dependence of the coefficients and an extra term $\varsigma_{nl}^{(in)}(\omega,q)P_4(\cos\theta)$. Even in this case expressions for $\beta_{nl}^{(in)}(\omega,q)$, $\gamma_{nl}^{(in)}(\omega,q)$, $\eta_{nl}^{(in)}(\omega,q)$, and $\varsigma_{nl}^{(in)}(\omega,q)$ via $g_{kl'L'}(q)$ are too complex as compared to relations for $\beta_{nl}(\omega)$, $\gamma_{nl}(\omega)$, and $\eta_{nl}(\omega)$ in photoionization. Therefore, it is more convenient to present the results for $s$, $p$, and $d$ subshells separately.

For $s$-subshells it is obtained

$$\frac{df_{n0}(q,\omega)}{d\Omega}=\sum_{L',L''=0}^{2}\frac{df_{n0}^{(L'L'')}(q,\omega)}{d\Omega}=\frac{f_{n0}(q,\omega)}{4\pi}\{1+$$
$$\frac{6}{w_0}g_{11}\left[g_{00}\cos(\delta_0-\delta_1)+2g_{22}\cos(\delta_1-\delta_2)\right]P_1(\cos\theta)+$$
$$\frac{2}{7w_0}\left[21g_{11}^2+5g_{22}(7g_{00}\cos(\delta_0-\delta_2)+5g_{22})\right]P_2(\cos\theta)+ \tag{14}$$
$$\frac{18}{w_0}g_{11}g_{22}\cos(\delta_1-\delta_2)P_3(\cos\theta)+\frac{90}{7W_0}g_{22}^2 P_4(\cos\theta)\}\equiv$$
$$\equiv\frac{f_{n0}(q,\omega)}{4\pi}\left\{1+\sum_{i=0}^{4}a_{n0i}(q,\omega)P_i(\cos\theta)\right\}.$$

where

$$f_{n0}=\frac{4\pi^2\omega}{(\omega-I_{n0})q^2}w_0;\ w_0=g_{00}^2+3g_{11}^2+5g_{22}^2. \tag{15}$$

Here and below for compactness of the expressions instead of $g_{klL}(q)$ we use $g_{lL}$.

For $l=1$ it is obtained

$$\frac{df_{n1}(q,\omega)}{d\Omega} = \sum_{L',L''=0}^{2} \frac{df_{n1}^{(L'L'')}(q,\omega)}{d\Omega} = \frac{f_{n1}}{4\pi}\{1+$$

$$\frac{1}{5w_1}[10g_{01}(2g_{12}-g_{10})\cos(\delta_0-\delta_1)+4g_{21}(5g_{10}-g_{12})\cos(\delta_1-\delta_2)+9g_{32}\cos(\delta_2-\delta_3)]P_1(\cos\theta)+$$

$$\frac{2}{7w_1}[7g_{21}(g_{21}-2g_{01}\cos(\delta_0-\delta_2))+7g_{12}(g_{12}-2g_{10})+$$

$$3g_{32}((7g_{10}-2g_{12})\cos(\delta_1-\delta_3)+4g_{32})]P_2(\cos\theta)+$$

$$\frac{6}{5w_1}[6g_{12}g_{21}\cos(\delta_1-\delta_2)+g_{32}(5g_{01}\cos(\delta_0-\delta_3)-4g_{21}\cos(\delta_2-\delta_3))]P_3(\cos\theta)+$$

$$\frac{18}{7w_1}g_{32}[g_{32}-4g_{12}\cos(\delta_1-\delta_3)]P_4(\cos\theta)\} \equiv \frac{f_{n1}(q,\omega)}{4\pi}\left\{1+\sum_{i=0}^{4}a_{n1i}(q,\omega)P_i(\cos\theta)\right\},$$

(16)

where

$$f_{n1} = \frac{4\pi^2\omega}{(\omega-I_{n1})q^2}w_1; \quad w_1 = g_{10}^2 + g_{01}^2 + 2\left[g_{21}^2 + g_{12}^2\right] + 3g_{32}^2.$$

(17)

For *l*=2 it is obtained

$$\frac{df_{n2}(q,\omega)}{d\Omega} = \frac{f_{n2}}{4\pi}\{1+\frac{6}{w_2}[14g_{11}(g_{22}-g_{20})-14g_{11}g_{02}\cos(\delta_0-\delta_1)+$$

$$3g_{31}((7g_{20}-2g_{22})\cos(\delta_2-\delta_3)+12g_{42}\cos(\delta_3-\delta_4))]P_1(\cos\theta)+$$

$$\frac{2}{245w_2}\Big[1029(g_{11}^2+6g_{31}^2)-18522g_{11}g_{31}\cos(\delta_1-\delta_3)+1225g_{02}(7g_{20}-10g_{22})\cos(\delta_0-\delta_2)-$$

$$125g_{22}(98g_{20}+15g_{22})+450g_{42}((49g_{20}-20g_{22})\cos(\delta_2-\delta_4)+25g_{42})\Big]P_2(\cos\theta)+$$

$$\frac{18}{w_2}[2g_{11}(g_{22}\cos(\delta_1-\delta_2)-6g_{42}\cos(\delta_1-\delta_4))+g_{31}(7g_{02}\cos(\delta_0-\delta_3)-$$

$$8g_{22}\cos(\delta_2-\delta_3)+6g_{42}\cos(\delta_3-\delta_4))]P_3(\cos\theta)+$$

$$\frac{90}{49w_2}\Big[20g_{22}^2+g_{42}(98g_{02}\cos(\delta_0-\delta_4)-100g_{22}\cos(\delta_2-\delta_4)+27g_{42})\Big]P_4(\cos\theta)\} \equiv$$

$$\equiv \frac{f_{n2}(q,\omega)}{4\pi}\left\{1+\sum_{i=0}^{4}a_{n2i}(q,\omega)P_i(\cos\theta)\right\},$$

(18)

where

$$f_{n2} = \frac{4\pi^2\omega}{35(\omega-I_{n2})q^2}w_2; \quad w_2 = 35g_{20}^2 + 42g_{11}^2 + 63g_{31}^2 + 35g_{02}^2 + 50g_{22}^2 + 90g_{42}^2.$$

(19)

Let us compare the result obtained in the small *q* limit with the known formula for photoionization of an atom by non-polarized light. To do this, we have to use the lowest order terms of the first three spherical Bessel functions:

$$j_0(qr) \cong 1 - \frac{(qr)^2}{6}; \quad j_1(qr) \cong \frac{qr}{3}\left(1 - \frac{(qr)^2}{10}\right); \quad j_2(qr) \cong \frac{(qr)^2}{15}\left(1 - \frac{(qr)^2}{14}\right). \tag{20}$$

It follows from (20) that the lowest in powers of $q$ term is $g_{11} \sim q \ll 1$[2]. Correction to $g_{11}$ is proportional to $q^3$. As to $g_{00}$ and $g_{22}$, they are proportional to $q^2$ with corrections of the order of $q^4$. Let us neglect in (14) terms of the order of $q^2$ and higher. Then one obtains the following expression:

$$\frac{df_{n0}(q,\omega)}{d\Omega} = \frac{f_{n0}(q,\omega)}{4\pi}\{1 + 2P_2(\cos\theta) + \frac{2g_{00}}{g_{11}}\cos(\delta_0 - \delta_1)P_1(\cos\theta) + \frac{2g_{22}}{g_{11}}\cos(\delta_1 - \delta_2)[2P_1(\cos\theta) + 3P_3(\cos\theta)]\}, \tag{21}$$

One should compare this relation with the similar one for photoionization of $n0$ subshell [15]:

$$\frac{d\sigma_{n0}(\omega)}{d\Omega} = \frac{\sigma_{n0}(\omega)}{4\pi}\{1 - P_2(\cos\theta) + \frac{\omega}{c}\frac{6q_2}{5d_1}\cos(\delta_1 - \delta_2)[P_1(\cos\theta) - P_3(\cos\theta)]\}. \tag{22}$$

According to (13), there are simple relations in $q \to 0$ limit between dipole $d_1$ and quadrupole $q_2$ matrix elements and $g_{11}$ and $g_{22}$: $g_{11} = qd_1/3$ and $g_{22} = 2q^2q_2/15$. With the help of relation $g_{00} = -q^2q_2/3$, (18) is transformed into the following expression:

$$\frac{df_{n0}(q,\omega)}{d\Omega} = \frac{f_{n0}(q,\omega)}{4\pi}\{1 + 2P_2(\cos\theta) + q\frac{4q_2}{5d_1}\cos(\delta_1 - \delta_2)[2P_1(\cos\theta) + 3P_3(\cos\theta)] - q\frac{2q_2}{d_{11}}\cos(\delta_0 - \delta_1)P_1(\cos\theta)\}. \tag{23}$$

In order to obtain the respective values in RPAE, one has to substitute phases $\delta_l$ and matrix elements $g_{ij}$ by the corresponding RPAE values:

$$\bar{\delta}_l = \delta_l + \Delta_l \text{ and } G_{ij}(q) \equiv \bar{G}_{ij}(q)\exp(i\Delta_l), \tag{24}$$

determined by (11). Then, the coefficients $a_{n0i}(q,\omega)$, $a_{n1i}(q,\omega)$, $a_{n2i}(q,\omega)$ in (14), (16) and (18) are transformed into $A_{n0i}(q,\omega)$, $A_{n1i}(q,\omega)$ and $A_{n2i}(q,\omega)$, respectively.

For differential in the outgoing electron angle GOS density of $nl$ subshell $dF_{nl}(q,\omega)/d\Omega$ the following relation are valid in RPAE

---

[2] As is seen from (8), we have in mind such values of $q$ that it is $qR_{nl} \ll 1$, where $R_{nl}$ is the radius of the ionized subshell.

$$\frac{dF_{nl}(q,\omega)}{d\Omega} = \sum_{L'L"} \frac{dF_{nl}^{L',L"}(q,\omega)}{d\Omega} = \frac{\omega\pi}{(\omega-I_{nl})q^2} \sum_{L'L"} (2L'+1)(2L"+1)i^{L'-L"} \times$$

$$\sum_{l'=|L'-l|}^{L'+l} \sum_{l"=|L"-l|}^{L"+l} \tilde{G}_{kl'L'}(q)\tilde{G}_{kl"L"}(q)i^{l"-l'}(2l'+1)(2l"+1)e^{i(\bar{\delta}_{l'}-\bar{\delta}_{l"})} \begin{pmatrix} L' & l & l' \\ 0 & 0 & 0 \end{pmatrix}\begin{pmatrix} L" & l & L" \\ 0 & 0 & 0 \end{pmatrix} \quad (25)$$

$$\sum_{L=|l'-l"|}^{l'+l"} P_L(\cos\theta)(-1)^{L+l}(2L+1)\begin{pmatrix} l' & L & l" \\ 0 & 0 & 0 \end{pmatrix}\begin{pmatrix} L & L' & L" \\ 0 & 0 & 0 \end{pmatrix}\begin{Bmatrix} L & L' & l" \\ l & l" & l' \end{Bmatrix}.$$

The partial value of GOS $F_{nl}(q,\omega)$ in RPAE is obtained from (25) by integrating over $d\Omega$, leading to the following expressions:

$$F_{nl}(q,\omega) = \sum_{L'} F_{nl}^{L'}(q,\omega) = \frac{4\omega\pi^2}{(\omega-I_{nl})q^2}\sum_{L'}(2L'+1)\sum_{l'=|L'-l|}^{L'+l}[\tilde{G}_{kl'L'}(q)]^2(2l'+1)\begin{pmatrix} L' & l & l' \\ 0 & 0 & 0 \end{pmatrix}^2 \quad \text{a)},$$

$$F_{nl}^{(W)}(q,\omega) = \sum_{L'} F_{nl}^{(W)L'}(q,\omega) = \frac{4\omega\pi^2}{q^2}\sum_{L'}(2L'+1)\sum_{l'=|L'-l|}^{L'+l}[\tilde{G}_{kl'L'}(q)]^2(2l'+1)\begin{pmatrix} L' & l & l' \\ 0 & 0 & 0 \end{pmatrix}^2 \quad \text{b)}. \quad (26)$$

Note that at small $q$ the dipole contribution in *weighted* GOSes $F_{nl}^{(W)}(q,\omega) = (\omega-I_{nl})F_{nl}(q,\omega)$ dominates and $F_{nl}^{(W)}(q,\omega)$ is simply proportional to the photoionization cross-section $\sigma_{nl}(\omega)$ [10]. To compare the results obtained with known formulas for the photoionization with lowest order non-dipole corrections taken into account, let us consider so small $q$ that it is enough to take into account terms with $L', L" \leq 2$. In this case, GOS angular distribution (23) can be presented similar to the photoionization case as

$$\frac{dF_{nl}(q,\omega)}{d\Omega} = \frac{F_{nl}(q,\omega)}{4\pi}\left\{1-\frac{\beta_{nl}^{(in)}(\omega,q)}{2}P_2(\cos\theta)+q\left[\gamma_{nl}^{(in)}(\omega,q)P_1(\cos\theta)+\eta_{nl}^{(in)}(\omega,q)P_3(\cos\theta)+\right.\right.$$
$$\left.\left.\varsigma_{nl}^{(in)}(\omega,q)P_4(\cos\theta)\right]\right\}. \quad (27)$$

The obvious difference is the $q$ dependence of the coefficients and an extra term $\varsigma_{nl}^{(in)}(\omega,q)P_4(\cos\theta)$. Even in this case expressions for $\beta_{nl}^{(in)}(\omega,q)$, $\gamma_{nl}^{(in)}(\omega,q)$, $\eta_{nl}^{(in)}(\omega,q)$, and $\varsigma_{nl}^{(in)}(\omega,q)$ via $g_{kl'L'}(q)$ are too complex as compared to relations for $\beta_{nl}(\omega)$, $\gamma_{nl}(\omega)$, and $\eta_{nl}(\omega)$ in photoionization. Therefore, it is more convenient to present the results for $s$, $p$, and $d$ subshells separately.

For $s$-subshells it is obtained

$$\frac{dF_{n0}(q,\omega)}{d\Omega} = \sum_{L',L"=0}^{2}\frac{dF_{n0}^{L',L"}(q,\omega)}{d\Omega} = \frac{F_{n0}(q,\omega)}{4\pi}\{1+$$

$$\frac{6}{W_0}\tilde{G}_{11}\left[\tilde{G}_{00}\cos(\bar{\delta}_0-\bar{\delta}_1)+2\tilde{G}_{22}\cos(\bar{\delta}_1-\bar{\delta}_2)\right]P_1(\cos\theta)+$$

$$\frac{2}{7W_0}\left[21\tilde{G}_{11}^2+5\tilde{G}_{22}(7\tilde{G}_{00}\cos(\bar{\delta}_0-\bar{\delta}_2)+5\tilde{G}_{22})\right]P_2(\cos\theta)+ \quad , (28)$$

$$\frac{18}{W_0}\tilde{G}_{11}\tilde{G}_{22}\cos(\bar{\delta}_1-\bar{\delta}_2)P_3(\cos\theta)+\frac{90}{7W_0}\tilde{G}_{22}^2P_4(\cos\theta)\} \equiv \frac{F_{n0}(q,\omega)}{4\pi}\left\{1+\sum_{i=1}^{4}A_{n0i}(q,\omega)P_i(\cos\theta)\right\}$$

where

$$F_{n0} = \frac{4\pi^2 \omega}{(\omega - I_{n0})q^2} W_0; \quad W_0 = \tilde{G}_{00}^2 + 3\tilde{G}_{11}^2 + 5\tilde{G}_{22}^2 \qquad (29)$$

Here and below for compactness of the expressions we use $\tilde{G}_{lL}$ instead of $\tilde{G}_{R\,nl\rightarrow kl'}^L(q,\omega)$ from (11).

For $l=1$ it is obtained

$$\frac{dF_{n1}(q,\omega)}{d\Omega} = \sum_{L',L''=0}^{2} \frac{dF_{n1}^{L',L''}(q,\omega)}{d\Omega} = \frac{F_{n1}}{4\pi}\{1+$$

$$\frac{1}{5W_1}\Big[10\tilde{G}_{01}(2\tilde{G}_{12}-\tilde{G}_{10})\cos(\bar{\delta}_0-\bar{\delta}_1)+4\tilde{G}_{21}(5\tilde{G}_{10}-\tilde{G}_{12})\cos(\bar{\delta}_1-\bar{\delta}_2)+9\bar{G}_{32}\cos(\bar{\delta}_2-\bar{\delta}_3)\Big]P_1(\cos\theta)+$$

$$\frac{2}{7W_1}\Big[7\tilde{G}_{21}(\tilde{G}_{21}-2\tilde{G}_{01}\cos(\bar{\delta}_0-\bar{\delta}_2))+7\tilde{G}_{12}(\tilde{G}_{12}-2\tilde{G}_{10})+$$

$$3\tilde{G}_{32}9(7\tilde{G}_{10}-2\tilde{G}_{12})\cos(\bar{\delta}_1-\bar{\delta}_3)+4\tilde{G}_{32})\Big]P_2(\cos\theta)+$$

$$\frac{6}{5W_1}\Big[6\tilde{G}_{21}\tilde{G}_{12}\cos(\bar{\delta}_1-\bar{\delta}_2)+\tilde{G}_{32}(5\tilde{G}_{01}\cos(\bar{\delta}_0-\bar{\delta}_3)-4\tilde{G}_{21}\cos(\bar{\delta}_2-\bar{\delta}_3))\Big]P_3(\cos\theta)+$$

$$\frac{18}{7W_1}\tilde{G}_{32}\Big[\tilde{G}_{32}-4\tilde{G}_{12}\cos(\bar{\delta}_1-\bar{\delta}_3)\Big]P_4(\cos\theta)\} \equiv \frac{F_{n1}(q,\omega)}{4\pi}\{1+\sum_{i=1}^{4}A_{n1i}(q,\omega)P_i(\cos\theta)\}$$

(30)

where

$$F_{n1} = \frac{4\pi^2\omega}{(\omega-I_{n1})q^2}W_1; \quad W_1 = \tilde{G}_{10}^2 + \tilde{G}_{01}^2 + 2\Big[\tilde{G}_{21}^2 + \tilde{G}_{12}^2\Big] + 3\tilde{G}_{32}^2. \qquad (31)$$

For $l=2$ it is obtained

$$\frac{dF_{n2}(q,\omega)}{d\Omega} = \frac{F_{n2}}{4\pi}\{1+\frac{6}{W_2}\Big[14\tilde{G}_{11}(\tilde{G}_{22}-\tilde{G}_{20})-14\tilde{G}_{11}\tilde{G}_{02}\cos(\bar{\delta}_0-\bar{\delta}_1)+$$

$$3\tilde{G}_{31}\big((7\tilde{G}_{20}-2\tilde{G}_{22})\cos(\bar{\delta}_2-\bar{\delta}_3)+12\tilde{G}_{42}\cos(\bar{\delta}_3-\bar{\delta}_4)\big)\Big]P_1(\cos\theta)+$$

$$\frac{2}{245W_2}\Big[1029(\tilde{G}_{11}^2+6\tilde{G}_{31}^2)-18522\tilde{G}_{11}\tilde{G}_{31}\cos(\bar{\delta}_1-\bar{\delta}_3)+1225\tilde{G}_{02}(7\tilde{G}_{20}-10\tilde{G}_{22})\cos(\bar{\delta}_0-\bar{\delta}_2)-$$

$$125\tilde{G}_{22}(98\tilde{G}_{20}+15\tilde{G}_{22})+450\tilde{G}_{42}(49\tilde{G}_{20}-20\tilde{G}_{22})\cos(\bar{\delta}_2-\bar{\delta}_4)+25\tilde{G}_{42}^2\Big]P_2(\cos\theta)+$$

$$\frac{18}{W_2}\Big[2\tilde{G}_{11}(\tilde{G}_{22}\cos(\bar{\delta}_1-\bar{\delta}_2)-6\tilde{G}_{42}\cos(\bar{\delta}_1-\bar{\delta}_4))+\tilde{G}_{31}(7\tilde{G}_{02}\cos(\bar{\delta}_0-\bar{\delta}_3)-$$

$$-8\tilde{G}_{22}\cos(\bar{\delta}_2-\bar{\delta}_3)+6\tilde{G}_{42}\cos(\bar{\delta}_3-\bar{\delta}_4))\Big]P_3(\cos\theta)+$$

$$\frac{90}{49W_2}\Big[20\tilde{G}_{22}^2+\tilde{G}_{42}(98\tilde{G}_{02}\cos(\bar{\delta}_0-\bar{\delta}_4)-100\tilde{G}_{22}\cos(\bar{\delta}_2-\bar{\delta}_4)+27\tilde{G}_{42})\Big]P_4(\cos\theta)\} \equiv$$

$$\equiv \frac{F_{n2}(q,\omega)}{4\pi}\{1+\sum_{i=1}^{4}A_{n2i}(q,\omega)P_i(\cos\theta)\},$$

(32)

where

$$F_{n2} = \frac{4\pi^2 \omega}{35(\omega - I_{n2})Eq^2} W_2; \quad W_2 = 35\tilde{G}_{20}^2 + 42\tilde{G}_{11}^2 + 63\tilde{G}_{31}^2 + 35\tilde{G}_{02}^2 + 50\tilde{G}_{22}^2 + 90\tilde{G}_{42}^2. \qquad (33)$$

Thus, we have expressed the angular distribution of knocked-out electrons in fast projectile-atom scattering via anisotropy parameters $A_{nli}(q,\omega)$, both dipole and non-dipole.

Let us compare the result obtained in the small $q$ limit with the known formula for photoionization of an atom by non-polarized light. To do this, we have to use the lowest order terms of the first three spherical Bessel functions (20).

The lowest in powers of $q$ term is $\tilde{G}_{11} \sim q \ll 1$[3]. Correction to $\tilde{G}_{11}$ is proportional to $q^3$. As to $\tilde{G}_{00}$ and $\tilde{G}_{22}$, they are proportional to $q^2$ with corrections of the order of $q^4$. Retaining in (27) terms of the order of $q^2$ and bigger, one has the following expression:

$$\frac{dF_{n0}(q,\omega)}{d\Omega} = \frac{F_{n0}(q,\omega)}{4\pi}\left\{1 + 2P_2(\cos\theta) + \frac{2\tilde{G}_{00}}{\tilde{G}_{11}}\cos(\bar{\delta}_0 - \bar{\delta}_1)P_1(\cos\theta) + \right.$$
$$\left. \frac{2\tilde{G}_{22}}{\tilde{G}_{11}}\cos(\bar{\delta}_1 - \bar{\delta}_2)[2P_1(\cos\theta) + 3P_3(\cos\theta)]\right\} \equiv$$
$$\equiv \frac{F_{n0}(q,\omega)}{4\pi}\left\{1 + 2P_2(\cos\theta) + \frac{2}{\tilde{G}_{11}}\left[\tilde{G}_{00}\cos(\bar{\delta}_0 - \bar{\delta}_1) + 2\tilde{G}_{22}\cos(\bar{\delta}_1 - \bar{\delta}_2)\right]P_1(\cos\theta) + \right. \qquad (34)$$
$$\left. \frac{6\tilde{G}_{22}}{\tilde{G}_{11}}\cos(\bar{\delta}_1 - \bar{\delta}_2)P_3(\cos\theta)\right\} \equiv$$
$$\equiv \frac{F_{n0}(q,\omega)}{4\pi}\left\{1 + 2P_2(\cos\theta) + q\gamma_{n0}^{(in)}(q,\omega)P_1(\cos\theta) + q\eta_{n0}^{(in)}(q,\omega)P_3(\cos\theta)\right\}$$

(compare with (27))

One should compare this relation with the similar one for photoionization of $n0$ subshell that [17]:

$$\frac{d\sigma_{n0}(\omega)}{d\Omega} = \frac{\sigma_{n0}(\omega)}{4\pi}\left\{1 - P_2(\cos\theta) + \kappa\frac{6\tilde{Q}_2}{5\tilde{D}_1}\cos(\bar{\delta}_1 - \bar{\delta}_2)[P_1(\cos\theta) - P_3(\cos\theta)]\right\} \equiv$$
$$\equiv \frac{\sigma_{n0}(\omega)}{4\pi}\left\{1 - P_2(\cos\theta) + \kappa\gamma_{n0}(\omega)P_1(\cos\theta) + \kappa\eta_{n0}(\omega)P_3(\cos\theta)\right\}. \qquad (35)$$

where $\gamma_{n0}(\omega) = -\eta_{n0}(\omega) = \frac{6\tilde{Q}_2}{5\tilde{D}_1}\cos(\bar{\delta}_1 - \bar{\delta}_2)$.

The difference between (34) and (35) is seen in the sign and magnitude of the dipole parameters and in different expressions for the non-dipole.

The difference between angular distributions of knocked out atomic electrons in fast projectile-atom scattering (34) and photoelectrons (35) exist and is essential even in the so-called optical limit $q \to 0$. According to (20), there are simple relations in $q \to 0$ limit between dipole $\tilde{D}_1$ and quadrupole $\tilde{Q}_2$ matrix elements and $\tilde{G}_{11}$, $\tilde{G}_{22}$: $\tilde{G}_{11} = q\tilde{D}_1/3$ and $\tilde{G}_{22} = 2q^2\tilde{Q}_2/15$. With the help of relations $\tilde{G}_{00} = -q^2\tilde{Q}_2/3 = -(5/2)\tilde{G}_{22}$, (34) is transformed into the following expression:

---

[3] As is seen from (20), we have in mind such values of $q$ that it is $qR_{nl} < 1$, where $R_{nl}$ is the radius of the ionized subshell.

$$\frac{dF_{n0}(q,\omega)}{d\Omega} = \frac{F_{n0}(q,\omega)}{4\pi}\{1+2P_2(\cos\theta)+q\frac{4\tilde{Q}_2}{5\tilde{D}_1}\cos(\bar{\delta}_1-\bar{\delta}_2)[2P_1(\cos\theta)+3P_3(\cos\theta)]$$

$$-q\frac{2\tilde{Q}_2}{\tilde{D}_1}\cos(\bar{\delta}_0-\bar{\delta}_1)P_1(\cos\theta)\} = \frac{F_{n0}(q,\omega)}{4\pi}\{1+2P_2(\cos\theta)+q\frac{2}{3}\gamma_{n0}(\omega)[2P_1(\cos\theta)+3P_3(\cos\theta)]$$

$$-q\frac{2\tilde{Q}_2}{\tilde{D}_1}\cos(\bar{\delta}_0-\bar{\delta}_1)P_1(\cos\theta)\} \equiv$$

$$\frac{F_{n0}(q,\omega)}{4\pi}\left\{1+2P_2(\cos\theta)+q\frac{2\tilde{Q}_2}{\tilde{D}_1}\left[\frac{4}{5}\cos(\delta_1-\delta_2)-\cos(\delta_0-\delta_1)\right]P_1(\cos\theta)+2q\gamma_{n0}(\omega)P_3(\cos\theta)\right\}$$

(36)

The deviation from (35) is evident, since the angular distribution is not expressed via a single non-dipole parameter $\gamma_{n0}(\omega)$ - a new phase difference $\bar{\delta}_0 - \bar{\delta}_1$ appears. As a result, following relations appear at very small q:

$$\gamma_{n0}^{(in)}(\omega) = \frac{2\tilde{Q}_2}{\tilde{D}_1}\left[\frac{4}{5}\cos(\delta_1-\delta_2)-\cos(\delta_0-\delta_1)\right],$$

$$\eta_{n0}^{(in)}(\omega) = 2\gamma_{n0}(\omega) = \frac{12}{5}\frac{\tilde{Q}_2}{\tilde{D}_1}\cos(\delta_1-\delta_2).$$

(37)

For $l > 0$ even at very small q the relation between non-dipole parameters in photoionization and inelastic fast electron scattering are rather complex.

The similarity of general structure and considerable difference between (22) and (23) is evident. Indeed, the contribution of the non-dipole parameters can be enhanced, since the condition $\omega/c \ll q \ll R^{-1}$ is easy to achieve. Let us note that even neglecting the terms with $q$ (22) and (23) remain different: while in photoionization the angular distribution is proportional to $\sin^2\theta$ [see (22)], in inelastic scattering it is proportional to $\cos^2\theta$ [see (23)]. The reason for this difference is clear. In photoabsorption the atomic electron is "pushed" off the atom by the electric field of the photon, which is perpendicular to the direction of the light beam. In inelastic scattering the push comes along momentum $q$, so the preferential emission of the electrons along the $\vec{q}$ direction, i.e. the maximum at $\theta = 0$. Similar reason explains the difference in the non-dipole terms. Note that the last term due to monopole transition (23) is absent in photoabsorption angular distribution (23). It confirms that the angular distribution of GOS densities is richer than that of photoionization.

## 4. Calculation procedure

In order to obtain $df_{nl}(q,\omega)/d\Omega$ from experiment, one has to measure the yield of electrons emitted at a given angle $\theta$ with energy $\varepsilon = k^2/2 = \omega - I_{nl}$ in coincidence with the fast outdoing particle that looses energy $\omega$ and transfers to the target atom momentum $q$.

To calculate $df_{nl}(q,\omega)/d\Omega$ we use the numeric procedure described at length in [16]. Calculations are performed in the frame of Hartree-Fock and RPAE approximations. As a concrete object, we choose $3p^6$ and $3s^2$ subshells of Ar. This object is representative, demonstrating strong influence of electron correlations both for p and s-electrons.

Calculations are performed using equations (14), (16), (18) in HF and their respective modifications (25-28), (30). (32), (34-37) in RPAE, for the following values of $q$, $q = 0.1; 0.6; 1.1; 1.6; 2.1$, and $I_{3p,3s} < \omega < I_{3p,3s} + 5Ry$. The results are presented for $a_{nli}(q,\omega)$, $A_{nli}(q,\omega)$, differential in emission angle GOSes $dF_{nl}(q,\omega)/d\Omega$ and weighted differential in emission angle GOSes $dF_{nl}^{(W)}(q,\omega)/d\Omega$ in Fig.1-34. Much of the data are presented for the so called magic angle determined by relation $P_2(\cos\theta_m) = 0$. At this angle the pure dipole contribution is zero, so the non-dipole corrections are most prominent. The lowest value of $q$ corresponds to the photoionization limit, since $qR \ll 1$ and in the considered frequency range $\omega/c < 0.05 < q_{min} = 0.1$. The last inequality shows that we consider non-dipole corrections to GOSes that are much bigger than non-dipole corrections to photoionization.

## 5. Results of calculations

Fig. 1 depicts weighted differential in emission angle GOSes of knocked out electron in fast projectile- He atom collision in HF, given by (26, 27) at magic angle $P_2(\cos\theta_m) = 0$ and a set of q values $q$=0.1, 0.6, 1.1, 1.6, 2.1. GOSes form a maximum that for small q is similar, as it should be, to the photoionization cross-section (see in [10]). With $\omega$ growth one can see second and third maximums. With increase of q the GOS maximum decreases rather fast moving at the same time from 1s threshold to higher $\omega$.

Fig. 2 depicts weighted differential in emission angle GOSes of knocked out electron in fast projectile- Ar atom collision in HF, given by (26, 27) at magic angle $P_2(\cos\theta_m) = 0$ and a set of q values $q$=0.1, 0.6, 1.1, 1.6, 2.1. Outer 3p subshell is considered. GOSes form a maximum that for small q is similar, as it should be, to the photoionization cross-section (see in [10]). With $\omega$ growth one can see second and third maximums. With increase of q the first GOS maximum decreases rather fast, slightly moving to higher $\omega$. Other maximums remain almost unaffected.

Fig. 3 shows weighted differential in electron emission angle GOSes given by (26, 27) at magic angle $P_2(\cos\theta_m) = 0$ of Ar 3p-subshell in HF and RPAE for $q$=0.1 and $q$=1.1. As in Fig. 2, GOSes are similar, as it should be, to the photoionization cross-section (see in [10]). At q=1.1 the main GOS maximum is much smaller than for q=0.1. The minimum, that is similar to Cooper minimum in photoionization is particularly deep in RPAE at q=0.1.

Fig. 4 offers weighted differential in electron emission angle GOSes given by (26, 27) at magic angle $P_2(\cos\theta_m) = 0$ of 3s-subshell for Ar at $q$=0.1, 0.6, 1.1, 1.6, 2.1 in HF. They have maximums close to threshold and with increase of q acquire a structure more complex that at small q, where it is similar to photoionization cross-section given in [10], rapidly decrease with growth of $\omega$ forming a prominent minimum that becomes narrower and deeper with increase of q.

Fig. 5 depicts weighted differential in electron emission angle GOSes (26, 27) at magic angle $P_2(\cos\theta_m) = 0$ of 3s-subshell for Ar at q=0.1 and 1.1 in HF and RPAE. For q=0.1 the situation is similar to photoionization where RPAE brings in a correlation or interference minimum (see in [10]). For q=1.1 RPAE considerably decreases the height of the near threshold maximum.

Fig. 6 depicts weighted differential in electron emission angle GOSes of knocked out electron in fast projectile- Xe atom collision in HF, given by (26, 27) at magic angle $P_2(\cos\theta_m) = 0$ and a set of $q$ values $q$=0.1, 0.6, 1.1, 1.6, 2.1. Outer 5p subshell is considered. GOSes for small q are similar to photoionization cross-section (see in [10]) and their first maximum rapidly decreases with q growth, loosing its power. Additional maximums at higher $\omega$ are affected by increase of q not that effective.

Fig. 7 demonstrates the weighted differential in electron emission angle GOSes given by (26, 27) at magic angle $P_2(\cos\theta_m) = 0$ of 5p-subshell for Xe at q=0.1 and 1.1 in HF and RPAE. Role of RPAE correlations is quite impressive. GOSes curves in RPAE has several maximums at both q values.

Fig. 8 offers the weighted differential in electron emission angle GOSes given by (25) at magic angle $P_2(\cos\theta_m) = 0$ of 5s-subshell for Xe at $q=0.1, 0.6, 1.1, 1.6, 2.1$ in HF. GOSes for small q are similar to photoionization cross-section (see in [10]). Contrary to the situation for 5p, the maximum for q>0.1 is bigger than for q=0.1. With growth of q it increases and then starts to decrease.

Fig. 9 presents the weighted GOSes differential in emission angle of the knocked-out electron (26, 27) at magic angle $P_2(\cos\theta_m) = 0$ of 5s-subshell for Xe at q=0.1 and 1.1 in HF and RPAE. For small q the GOSes are similar to the photoionization cross-section (see in [10]). For q=1.1 the near threshold maximum become much stronger. One see a broad and low RPAE maximums at 15 and 12 Ry for q=0.1 and q=1.1, respectively.

Fig. 10 demonstrates the weighted differential in electron emission angle GOSes (26, 27) at magic angle $P_2(\cos\theta_m) = 0$ of 4d-subshell for Xe at $q=0.1, 0.6, 1.1, 1.6, 2.1$ in HF. GOSes have a powerful maximum, in general similar to the photoionization cross-section of 4d given in (see in [10]). With increase of q the maximums decrease and move to higher concentrate at smaller energies and a small maximum at q=0.1 become evident.

Fig. 11 presents the weighted GOSes differential in emission angle of the knocked-out electron (26, 27) at magic angle $P_2(\cos\theta_m) = 0$ of 4d-subshell for Xe at q=0.1 and 1.1 in HF and RPAE. GOSes have a powerful maximum in general similar to the photoionization cross-section of 4d given in (see in [10]). The RPAE role is not too impressive just as the influence of q growth from 0.1 to 0.01. This is natural in view of the smallness of the 4d-subshell radius as compared to 1.

Fig. 12 presents the angular anisotropy parameters of knocked-out electrons in fast projectile-atom collision $a_{1si}$ given by (28) as functions of $\omega$ at q=0.1 in HF for He. The dipole parameter $a_{1s2}$ is, as it should be for $q=0.1$, bigger than the non-dipole parameters $a_{1s1}$ $a_{1s3}$ by a factor of ten. Note that it is equal to 2 as it should be in the optical limit $q \to 0$ (see [10]). The limit $a_{1s4}$ is smaller than the dipole $a_{1s2}$ by two orders of magnitude.

Fig. 13 presents the angular anisotropy parameters of knocked-out electrons from 3p-subshell of Ar in fast projectile-atom collision, given by (28) in HF ($a_{3pi}$) and RPAE ($A_{3pi}$) at q=0.1 as functions of $\omega$. The role of correlations is essential only for $A_{3p1}$ that differs essentially from $a_{3p1}$. For other angular anisotropy parameters correlations are small and energy dependence is in the vicinity of threshold. The dipole coefficients, as it should be for small q, by far exceed the non-dipole.

Fig. 14 demonstrates the angular anisotropy parameters of knocked-out electrons from 3p-subshell of Ar in fast projectile-atom collision, given by (28), in HF ($a_{3pi}$) and RPAE ($A_{3pi}$) at $q=1.1$ as functions of $\omega$. The increase of q from 0.1 to 1.1 leads to essential growth of the non-dipole parameters with $i = 2$.

Fig. 15 shows the angular anisotropy parameters of knocked-out electrons from 3s-subshell of Ar in fast projectile-atom collision, given by (28), in HF ($a_{3si}$) and RPAE ($A_{3si}$) at $q=0.1$ as functions of $\omega$. Energy dependence of dipole parameter $a_{3s2}$ and $A_{3s2}$ disappear at about 4Ry and they reach the value of the optical limit ($q \to 0$) - $a_{3s2} = A_{3s2} = 2$. Parameters with $i = 1;3$ are by a factor of q=0.1 smaller than that with $i = 2$. As to parameter $i = 4$, it is smaller than $a_{3s2} = A_{3s2} = 2$ by a factor $q^2 = 0.01$.

Fig. 16 displays the angular anisotropy parameters of knocked-out electrons from 3p-subshell of Ar in fast projectile-atom collision, given by (28) in HF ($a_{3si}$) and RPAE ($A_{3si}$) at $q=1.1$ as functions of $\omega$. All parameters are of the same order of magnitude, essentially depend upon $\omega$ and that with $i=1$ and 4 are mirror reflection relative to horizontal axis of the parameter with $i=4$.

Fig. 17 presents the angular anisotropy parameters of knocked-out electrons in fast projectile - Xe atom collision of 5p-subshell, given by (28) in HF ($a_{5pi}$) and RPAE ($A_{5pi}$) at q=0.1 as functions of $\omega$. Note that dipole parameters $a_{5p2}$ and $A_{5p2}$ are close to zero, as well as $a_{5p1}$, $A_{5p1}$ and $a_{5p4}$, $A_{5p4}$. Very big is only the oscillating value of $a_{5p1}$ and $A_{5p1}$.

Fig. 18 gives the angular anisotropy parameters of knocked-out electrons in fast projectile-atom collision given by (28), of 5p-subshell on Xe in HF ($a_{3pi}$) and RPAE ($A_{3pi}$) at q=1.1 as functions of $\omega$. Note that dipole parameters $a_{5p2}$ and $A_{5p2}$, as well as $a_{5p1}$, $A_{5p1}$ and $a_{5p4}$, $A_{5p3}$ are considerably bigger than at q=0.1. As to $a_{5p1}$ and $A_{5p1}$, it becomes smaller.

Fig. 19 shows the angular anisotropy parameters of knocked-out electrons in fast projectile-atom collision given by (28) of 5s-subshell on Xe in HF ($a_{5si}$) and RPAE ($A_{5pi}$) at q=0.1 as functions of $\omega$. Note that the dipole parameters $a_{5s2}$ and $A_{5s2}$ are everywhere, except vicinity of 5s threshold and 4d Giant resonance, equal to 2, as it should be in the optical limit $q \to 0$. Near 5s threshold and in the 4d Giant resonance region other parameters essentially changes, whereas in other $\omega$ region they are smooth functions of $\omega$. The ratio between parameters is natural: $100 a_{5s4}, A_{5s4} \approx 10 a_{5s1}, A_{5s1}; a_{5s3}, A_{5s3} \approx a_{5s2}, A_{5s2}$, in accord with q=0.1 and $q^2 = 0.01$.

Fig. 20 presents the angular anisotropy parameters of knocked-out electrons in fast projectile-atom collision given by (28) of 5s-subshell on Xe in HF ($a_{5si}$) and RPAE ($A_{5pi}$) at q=1.1 as functions of $\omega$. All parameters are of the same order of magnitude and rapidly oscillate that is natural for big q.

Fig. 21 shows the angular anisotropy parameters of knocked-out electrons in fast projectile-atom collision as functions of $\omega$ given by (28) of 4d-subshell on Xe in HF ($a_{4di}$) and RPAE ($A_{4di}$) at q=0.1 as functions of $\omega$. The dipole parameter $i=2$ is of the order of one having a big and broad maximum at threshold. The difference between HF and RPAE is noticeable. All other parameters, except that with $i=1$ are very close to zero.

Fig. 22 demonstrates the angular anisotropy parameters of knocked-out electrons in fast projectile-atom collision given by (28) of 4d-subshell on Xe in HF ($a_{4di}$) and RPAE ($A_{4di}$) at q=1.1 as functions of $\omega$. As it is for q=0.1, here absolutely dominates the dipole parameter, but contrary to the case of q=0.1 parameter $i=1$ is also big. Noticeable is the parameter with $i=3$. As to $i=4$, this parameter is very small, almost zero.

Fig. 23 shows the angular anisotropy non-dipole parameters of knocked-out electrons in fast projectile-atom collision in the optical limit $\gamma_{1s}^{(in)}(\omega)$ and $\eta_{1s}^{(in)}(\omega)$ given by (36, 37) at $q$=0.1 compared to similar parameters in photoionization $\gamma_{1s}(\omega)$ and $\eta_{1s}(\omega)$, given by (35) on He in HF. The ratio $\eta_{ns}^{(in)}(\omega) = 2\gamma_{ns}(\omega)$ is fulfilled with good accuracy.

Fig. 24 demonstrates the angular anisotropy non-dipole parameters of knocked-out electrons in fast projectile-atom collision in the optical limit $\gamma_{1s}^{(in)}(\omega)$ and $\eta_{1s}^{(in)}(\omega)$ given by (36, 37) at $q$=0.2 compared to similar parameters in photoionization $\gamma_{1s}(\omega)$ and $\eta_{1s}(\omega)$, given by (35) on He in HF. The ratio $\eta_{ns}^{(in)}(\omega) = 2\gamma_{ns}(\omega)$ starts to be violated due to growth of $\eta_{1s}^{(in)}(q,\omega)$.

Fig. 25 presents the angular anisotropy non-dipole parameters of knocked-out electrons in fast projectile-atom collision in the optical limit given by (36, 37) at $q$=0.1 for 3s subshell of Ar. Strong variations of parameters are located at $\omega < 2Ry$. A strong and deep RPAE minimum in $\eta^{(in)}$ should be compared to much smaller maximum in RPAE only for $\gamma^{(in)}$ at the same $\omega$.

Fig. 26 shows the angular anisotropy non-dipole parameters of knocked-out electrons in fast projectile-atom collision in the optical limit $\gamma_{3s}^{(in)}(\omega)$ and $\eta_{3s}^{(in)}(\omega)$, given by (36, 37) at $q$=0.1 and compared to similar parameters in photoionization $\gamma_{3s}(\omega)$ and $\eta_{3s}(\omega)$, given by (35) for 3s subshell of Ar in RPAE. The relation $\eta_{ns}^{(in)}(\omega) = 2\gamma_{ns}(\omega)$ is fulfilled with good accuracy

Fig. 27 depicts the angular anisotropy non-dipole parameters of knocked-out electrons in fast projectile-atom collision in the optical limit, given by (36, 37) at q=1.1 for 3s subshell of Ar. Both parameters $\gamma^{(in)}$ and $\eta^{(in)}$ vary essentially and are of the same order of magnitude.

Fig. 28 compares the non-dipole anisotropy parameters of Ar 3s electrons in photoionization and fast electron inelastic scattering (see (36) and (37), respectively) at q=0, calculated in HF.

Fig. 29 confronts the non-dipole anisotropy parameters of Ar 3s electrons in photoionization and fast electron inelastic scattering (see (36) and (37), respectively) at q=0, calculated in RPAE.

Fig. 30 presents the angular anisotropy non-dipole parameters of knocked-out electrons in fast projectile-atom collision in the optical limit given by (36, 37) at q=0.1 for 5s subshell of Xe. Strong variations of parameters are located at $2 < \omega < 4 Ry$ and at $8 < \omega < 12 Ry$, i.e. near threshold and in the 4d Giant resonance vicinity. The variation at $8 < \omega < 12 Ry$ is a direct consequence of the action of 4d Giant resonance upon non-dipole parameters of 5s Xe. A strong and deep minimum in $\eta^{(in)}$ should be compared to a much smaller maximum in RPAE only for $\gamma^{(in)}$ at the same $\omega$.

Fig. 31 depicts the angular anisotropy non-dipole parameters of knocked-out electrons in fast projectile-atom collision in the optical limit given by (36, 37) at q=1.1 for 5s subshell of Xe. Both parameters $\gamma^{(in)}$ and $\eta^{(in)}$ are essentially varying functions, and of the same order of magnitude. This signals that q=1.1 is far from the optical limit.

Fig. 32 shows the angular anisotropy non-dipole parameters of knocked-out electrons in fast projectile-atom collision in the optical limit $\gamma_{5s}^{(in)}(\omega)$ and $\eta_{5s}^{(in)}(\omega)$, given by (36, 37) at $q$=0.1 compared to similar parameters in photoionization $\gamma_{5s}(\omega)$ and $\eta_{5s}(\omega)$, given by (35) for 5s subshell of Xe in RPAE. The ratio $\eta_{ns}^{(in)}(\omega) = 2\gamma_{ns}(\omega)$ is fulfilled with good accuracy

Fig. 33 compares the non-dipole anisotropy parameters of Xe 5s electrons in photoionization and fast electron inelastic scattering (see (36) and (37), respectively) at q=0, calculated in HF. Note that the relations that connect non-dipole angular anisotropy parameters presented in (35) and (37) are fulfilled accurately enough.

Fig 34 confronts the non-dipole anisotropy parameters of Xe 5s electrons in photoionization and fast electron inelastic scattering (see (36) and (37), respectively) at q=0, calculated in RPAE. Note that the relations that connect non-dipole angular anisotropy parameters presented in (35) and (37) are fulfilled accurately enough in RPAE just as in HF.

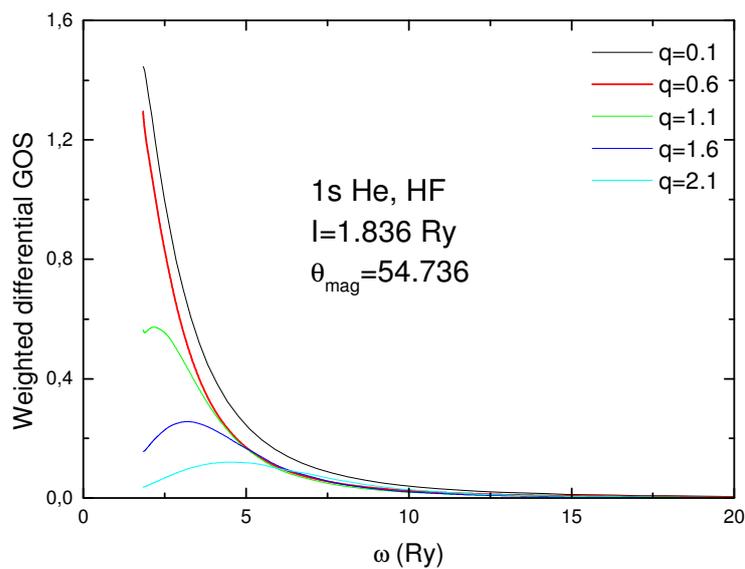

Fig. 1. Weighted differential generalized oscillator strength in HF given by (26, 27) at magic angle $P_2(\cos\theta_m) = 0$ of He at different q.

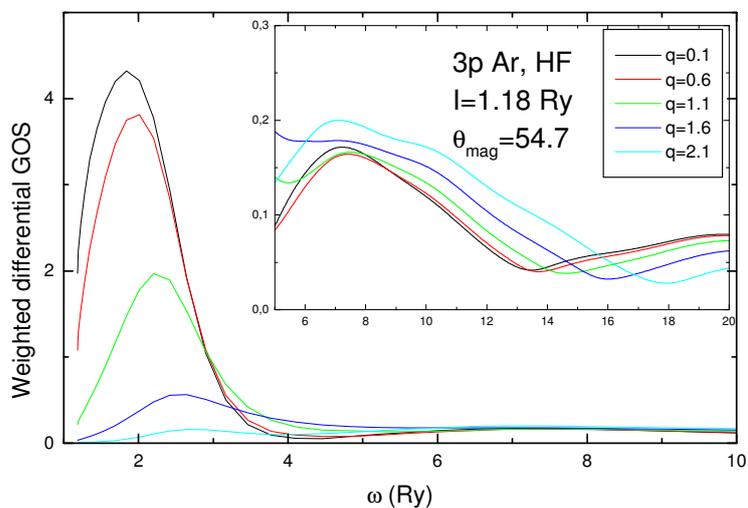

Fig. 2. Weighted differential generalized oscillator strength given by (26, 27) at magic angle $P_2(\cos\theta_m) = 0$ of 3p-subshell for Ar at different q in HF.

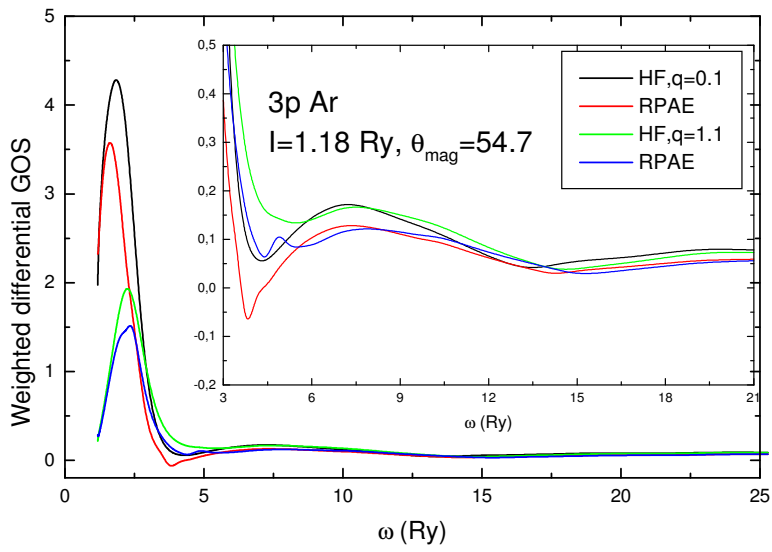

Fig. 3. Weighted differential generalized oscillator strength given by (26, 27) at magic angle $P_2(\cos\theta_m) = 0$ of 3p-subshell for Ar at different q in HF and RPAE.

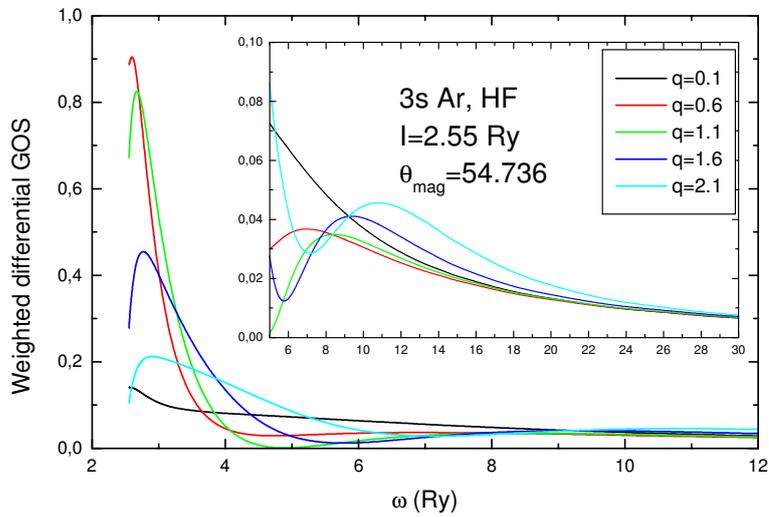

Fig. 4. Weighted differential generalized oscillator strength given by (26, 27) at magic angle $P_2(\cos\theta_m) = 0$ of 3s-subshell for Ar at different q in HF.

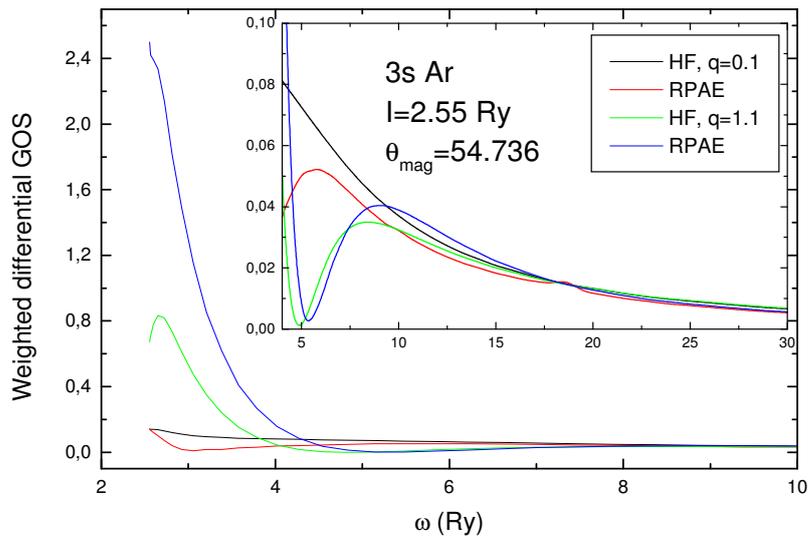

Fig. 5. Weighted differential generalized oscillator strength (26. 27) at magic angle $P_2(\cos\theta_m) = 0$ of 3s-subshell for Ar at different q in HF and RPAE.

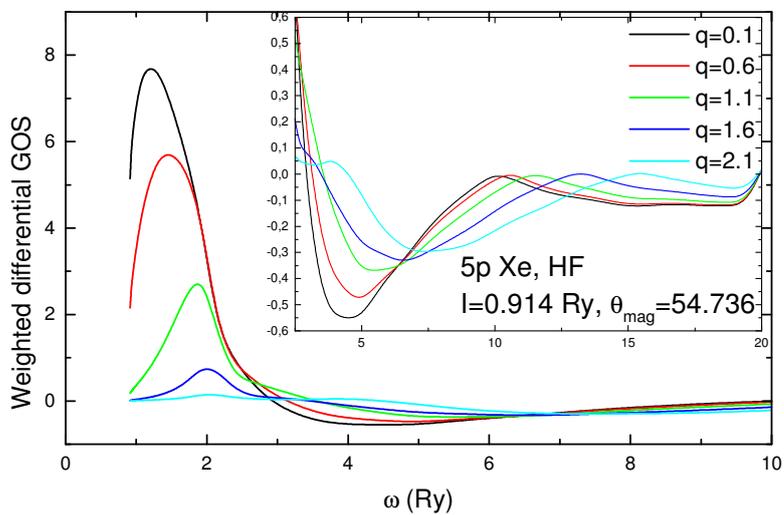

Fig. 6. Weighted differential generalized oscillator strength (26, 27) at magic angle $P_2(\cos\theta_m) = 0$ of 5p-subshell for Xe at different q in HF.

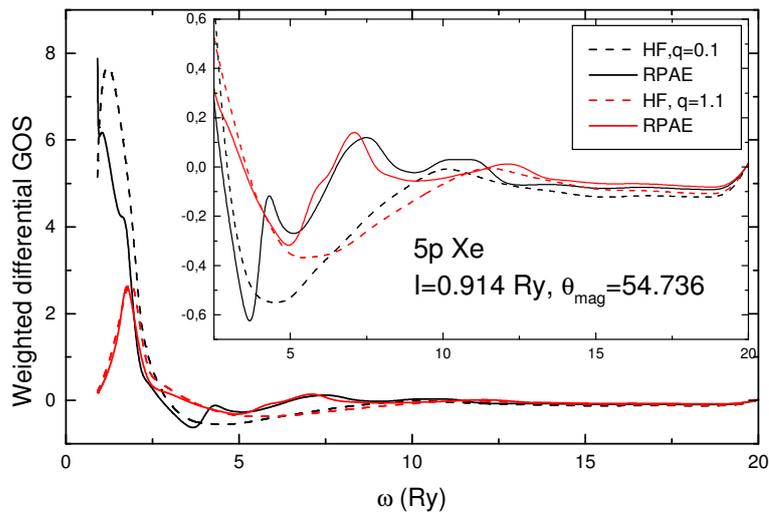

Fig. 7. Weighted differential generalized oscillator strength (26, 27) at magic angle $P_2(\cos\theta_m)=0$ of 5p-subshell for Xe at different q in HF and RPAE.

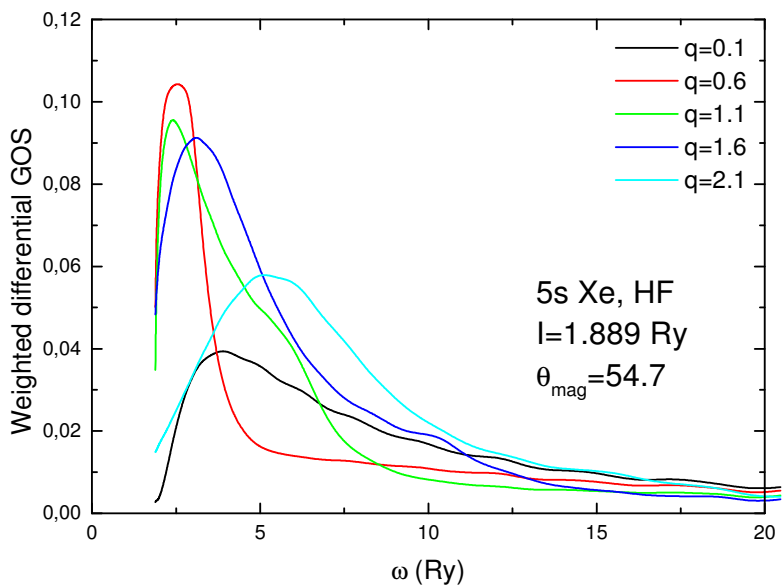

Fig. 8. Weighted differential generalized oscillator strength (26, 27) at magic angle $P_2(\cos\theta_m)=0$ of 5s-subshell for Xe at different q in

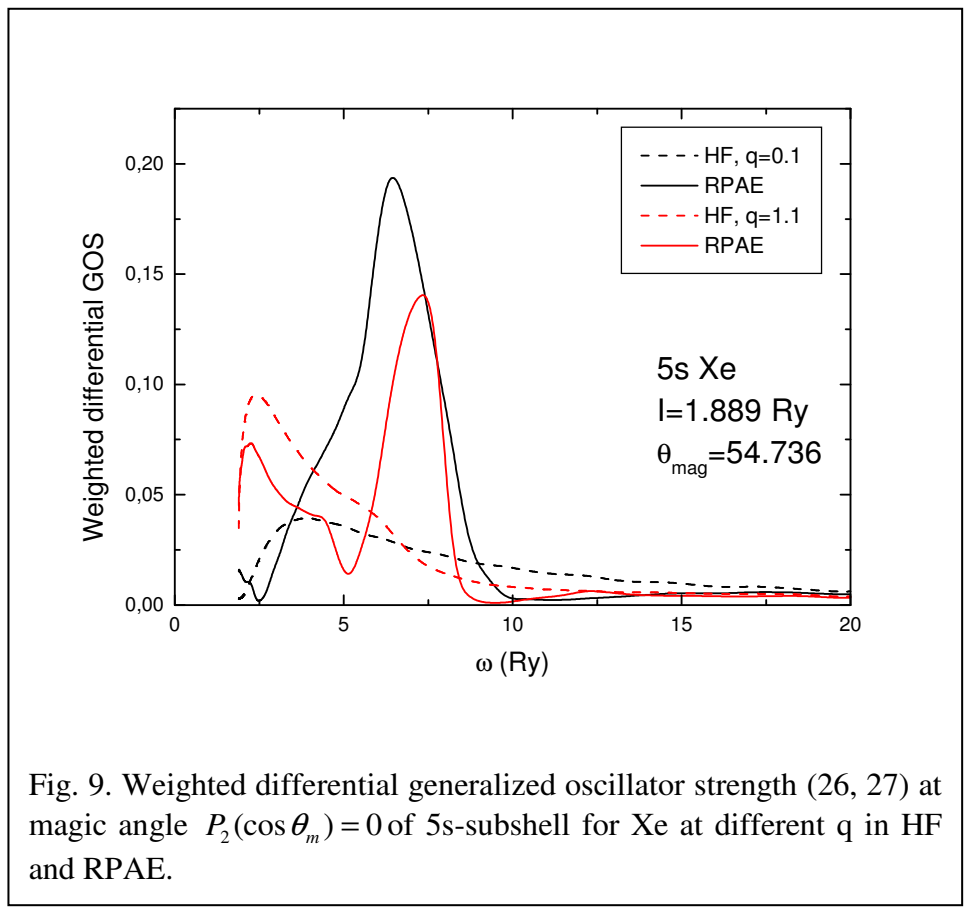

Fig. 9. Weighted differential generalized oscillator strength (26, 27) at magic angle $P_2(\cos\theta_m)=0$ of 5s-subshell for Xe at different q in HF and RPAE.

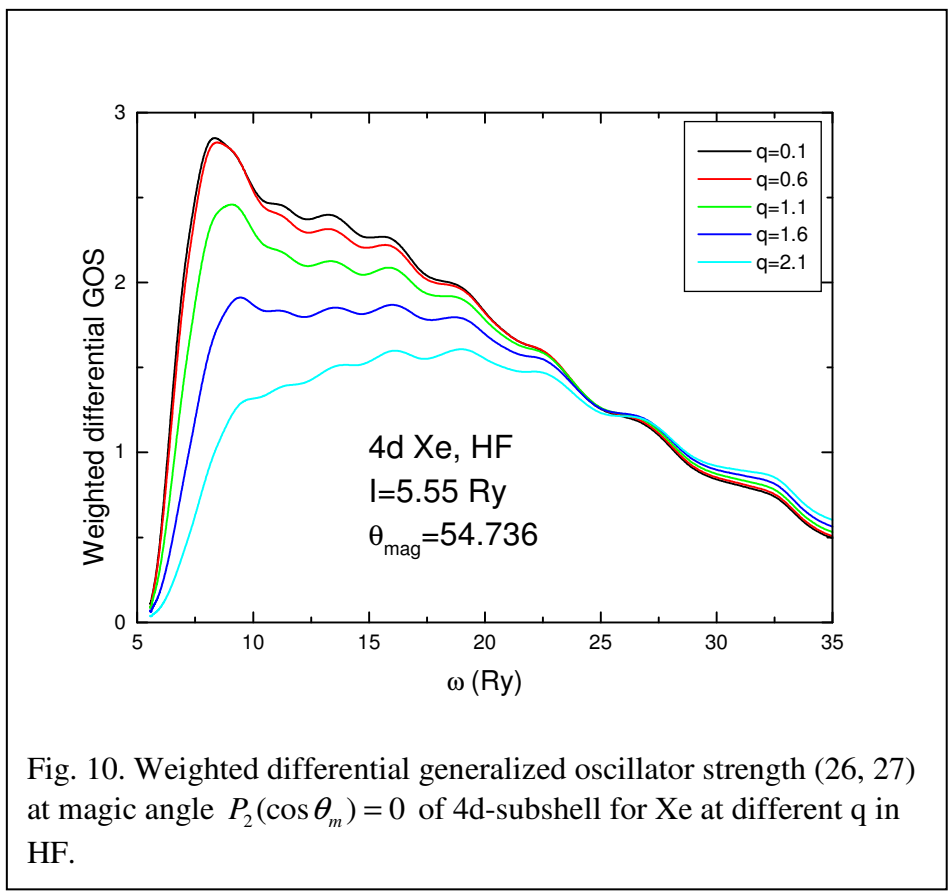

Fig. 10. Weighted differential generalized oscillator strength (26, 27) at magic angle $P_2(\cos\theta_m)=0$ of 4d-subshell for Xe at different q in HF.

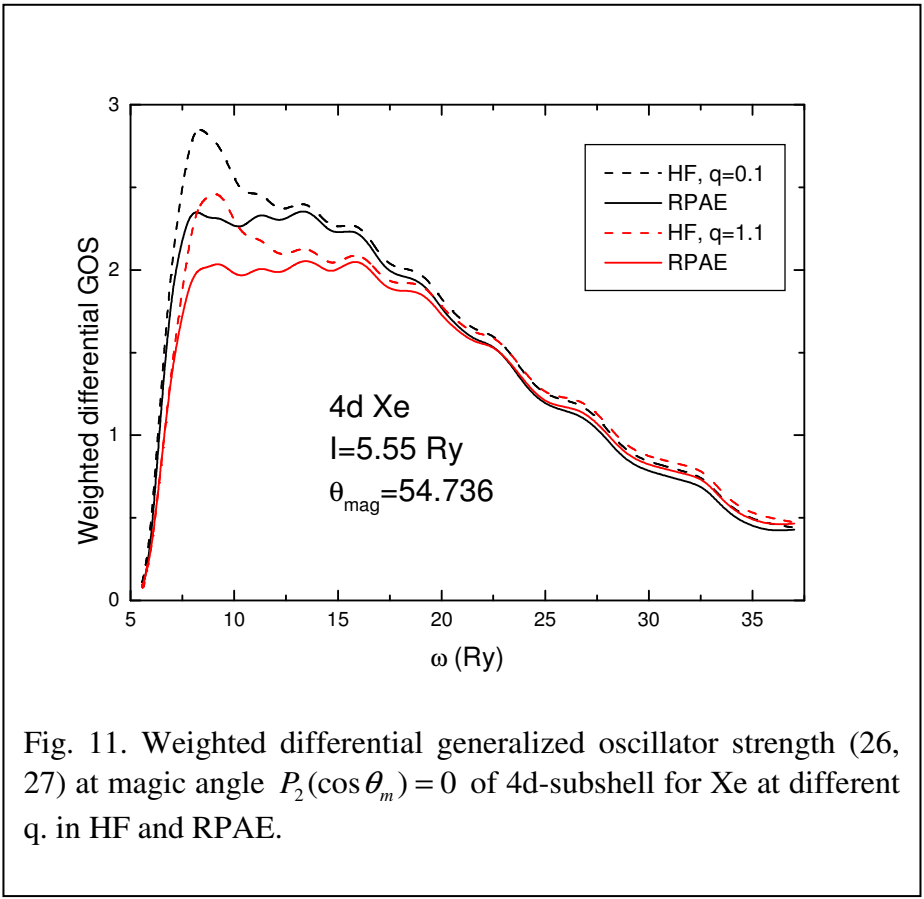

Fig. 11. Weighted differential generalized oscillator strength (26, 27) at magic angle $P_2(\cos\theta_m) = 0$ of 4d-subshell for Xe at different q. in HF and RPAE.

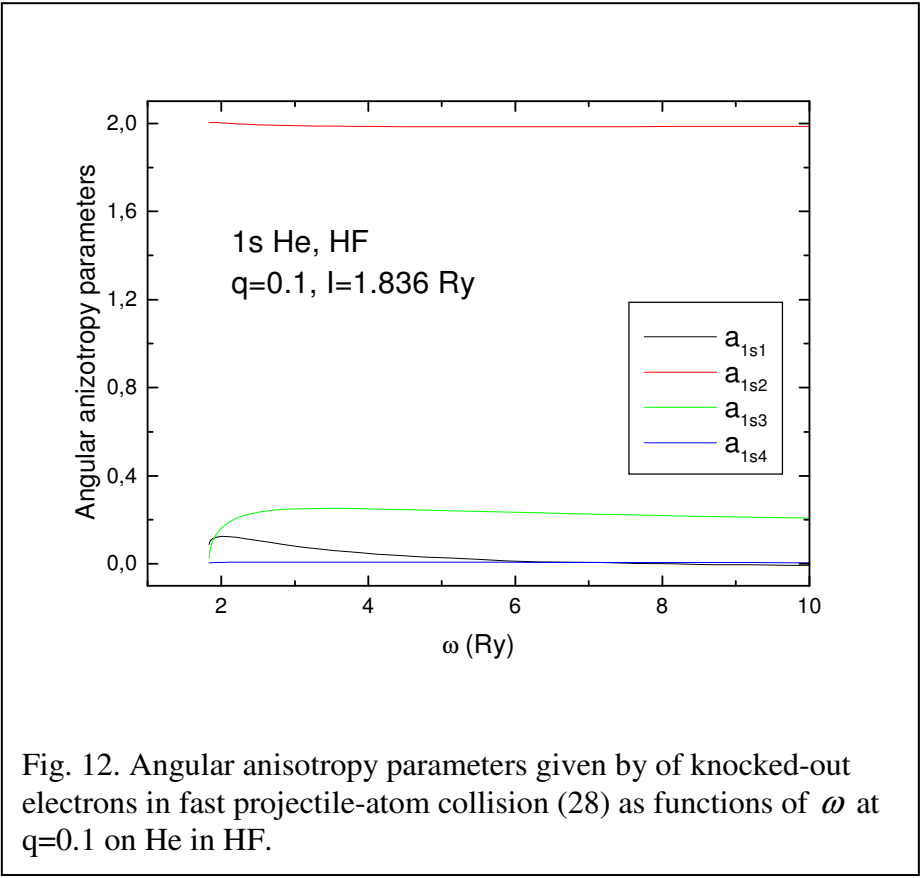

Fig. 12. Angular anisotropy parameters given by of knocked-out electrons in fast projectile-atom collision (28) as functions of $\omega$ at q=0.1 on He in HF.

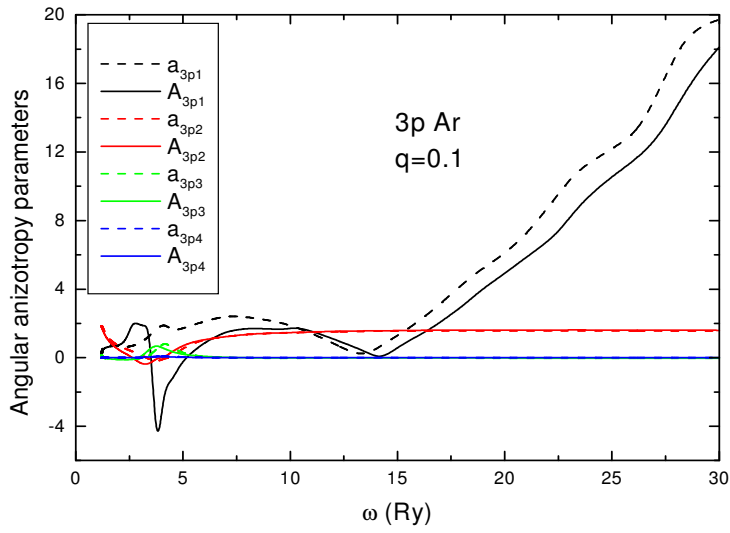

Fig. 13. Angular anisotropy parameters given by of knocked-out electrons in fast projectile-atom collision (28) of 3p-subshell on Ar in HF ($a_{3pi}$) and RPAE ($A_{3pi}$) at q=0.1 as functions of $\omega$.

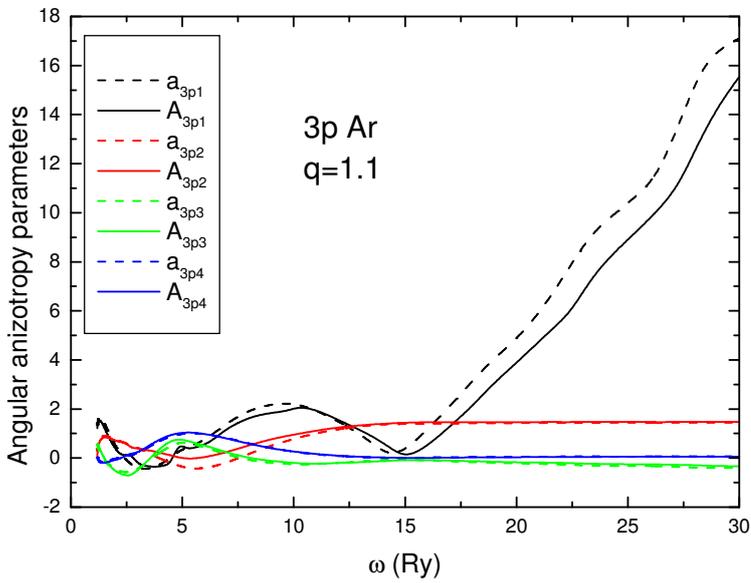

Fig. 14. Angular anisotropy parameters given by of knocked-out electrons in fast projectile-atom collision (28) as functions of $\omega$ of 3p-subshell on Ar in HF ($a_{3pi}$) and RPAE ($A_{3pi}$) at q=1.1.

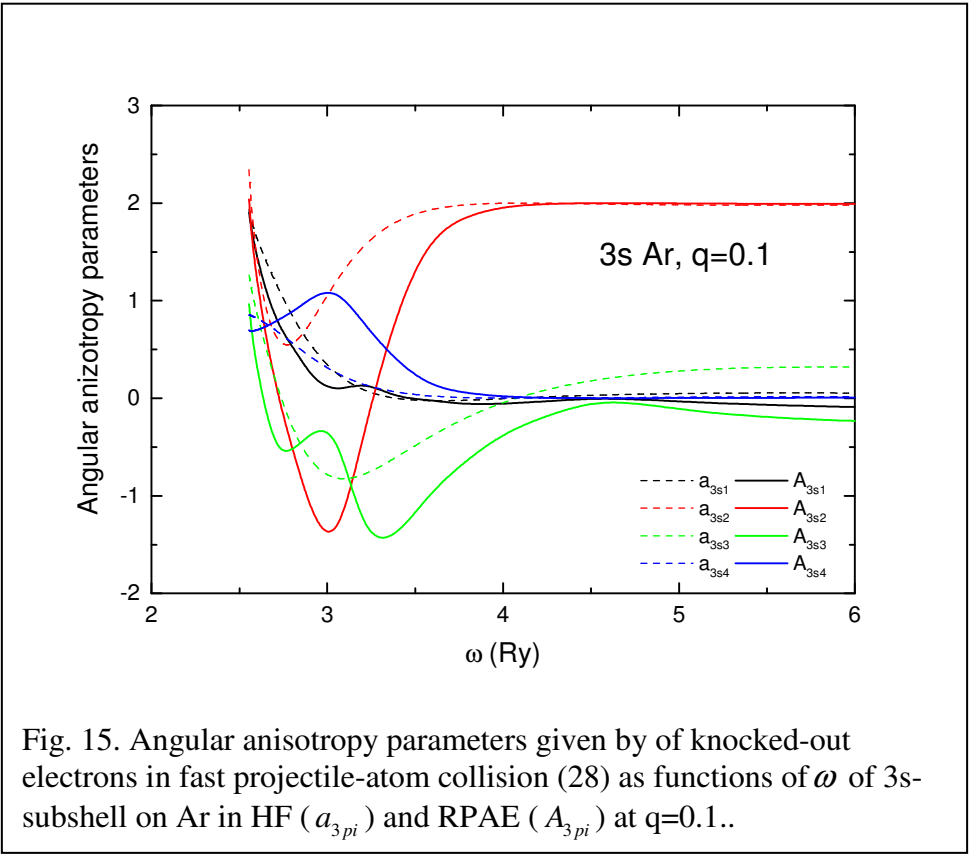

Fig. 15. Angular anisotropy parameters given by of knocked-out electrons in fast projectile-atom collision (28) as functions of $\omega$ of 3s-subshell on Ar in HF ($a_{3pi}$) and RPAE ($A_{3pi}$) at q=0.1..

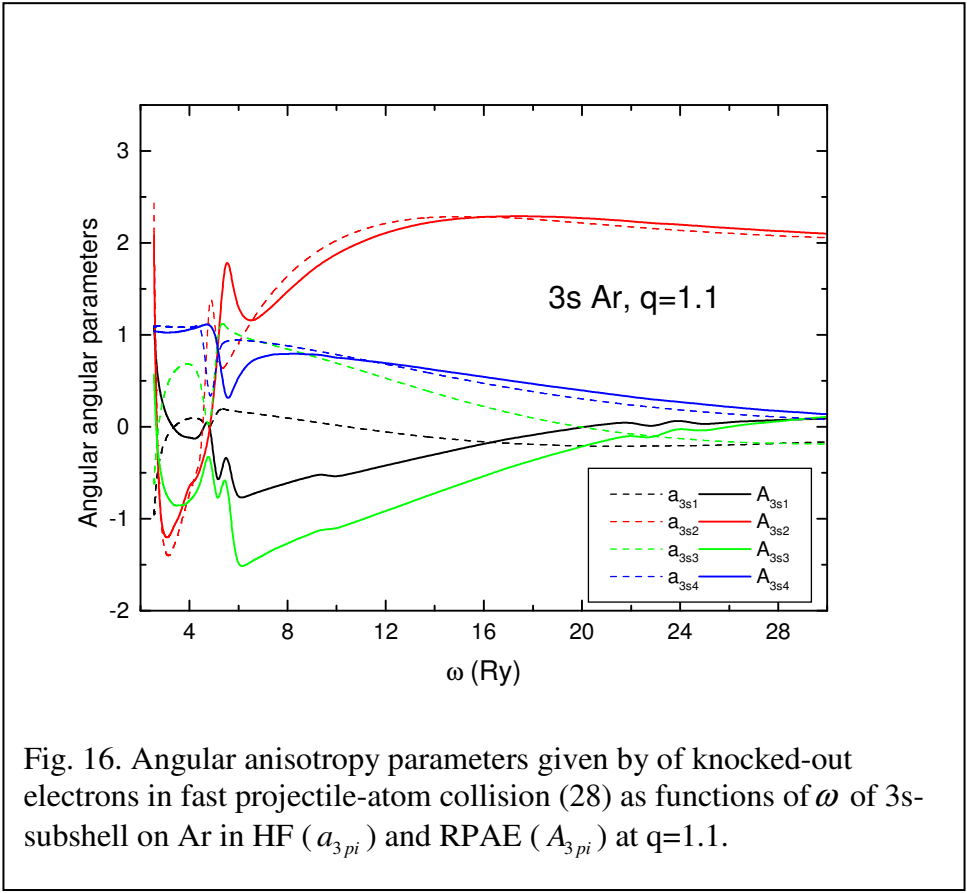

Fig. 16. Angular anisotropy parameters given by of knocked-out electrons in fast projectile-atom collision (28) as functions of $\omega$ of 3s-subshell on Ar in HF ($a_{3pi}$) and RPAE ($A_{3pi}$) at q=1.1.

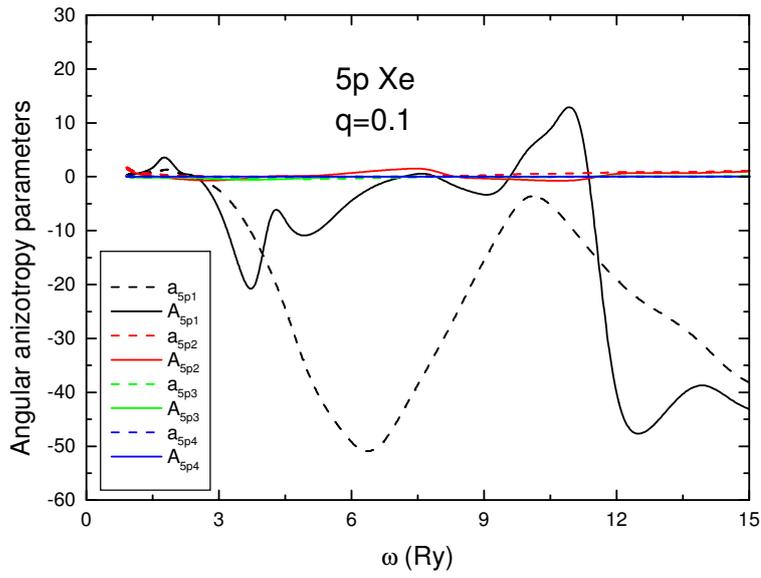

Fig. 17. Angular anisotropy parameters given by of knocked-out electrons in fast projectile-atom collision as functions of $\omega$ (28) of 5p-subshell on Xe in HF ($a_{3pi}$) and RPAE ($A_{3pi}$) at q=0.1.

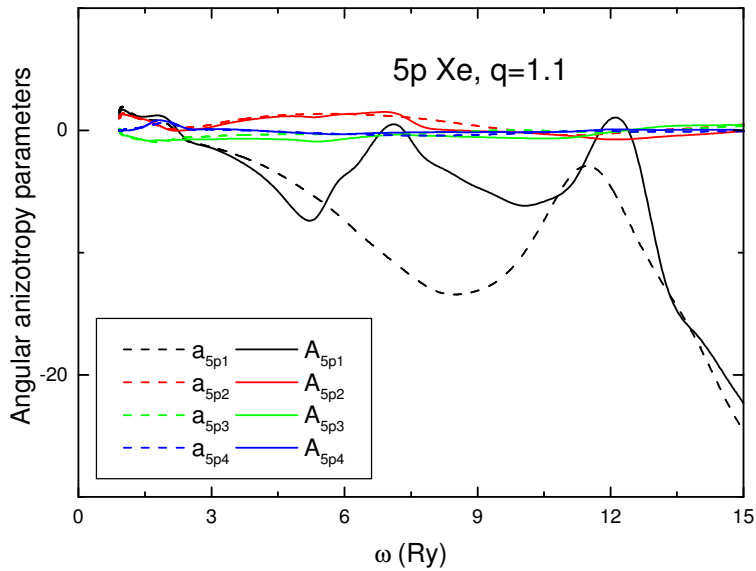

Fig. 18. Angular anisotropy parameters given by of knocked-out electrons in fast projectile-atom collision as functions of $\omega$ (28) of 5p-subshell on Xe in HF ($a_{3pi}$) and RPAE ($A_{3pi}$) at q=1.1.

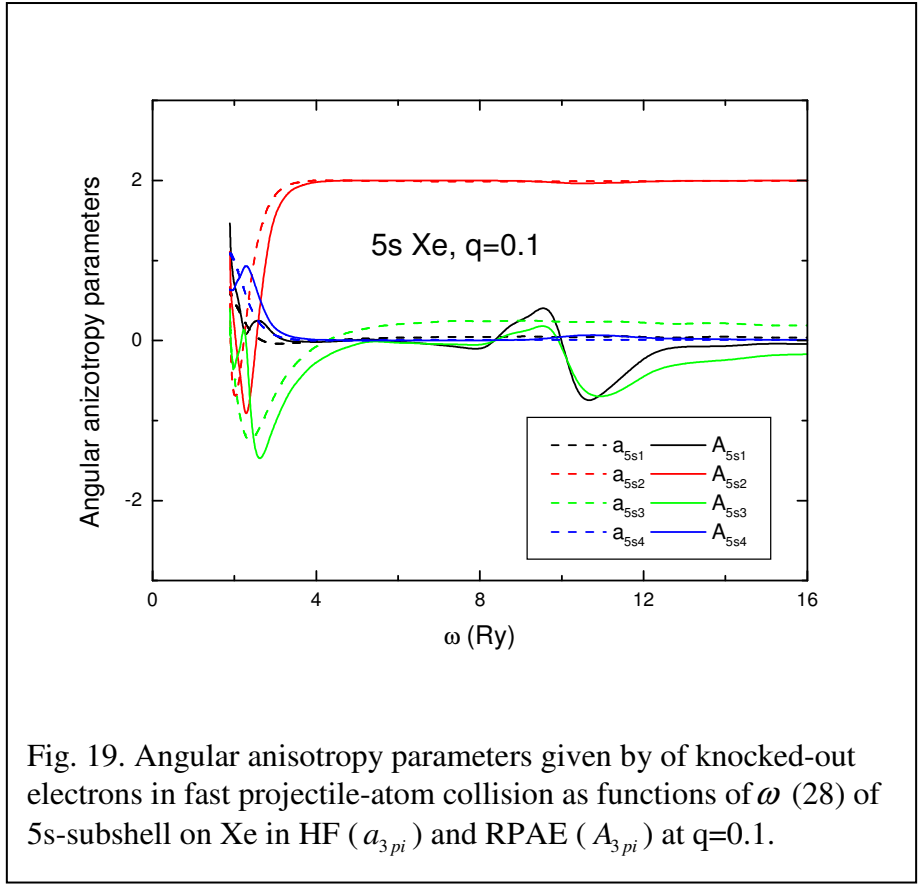

Fig. 19. Angular anisotropy parameters given by of knocked-out electrons in fast projectile-atom collision as functions of $\omega$ (28) of 5s-subshell on Xe in HF ($a_{3pi}$) and RPAE ($A_{3pi}$) at q=0.1.

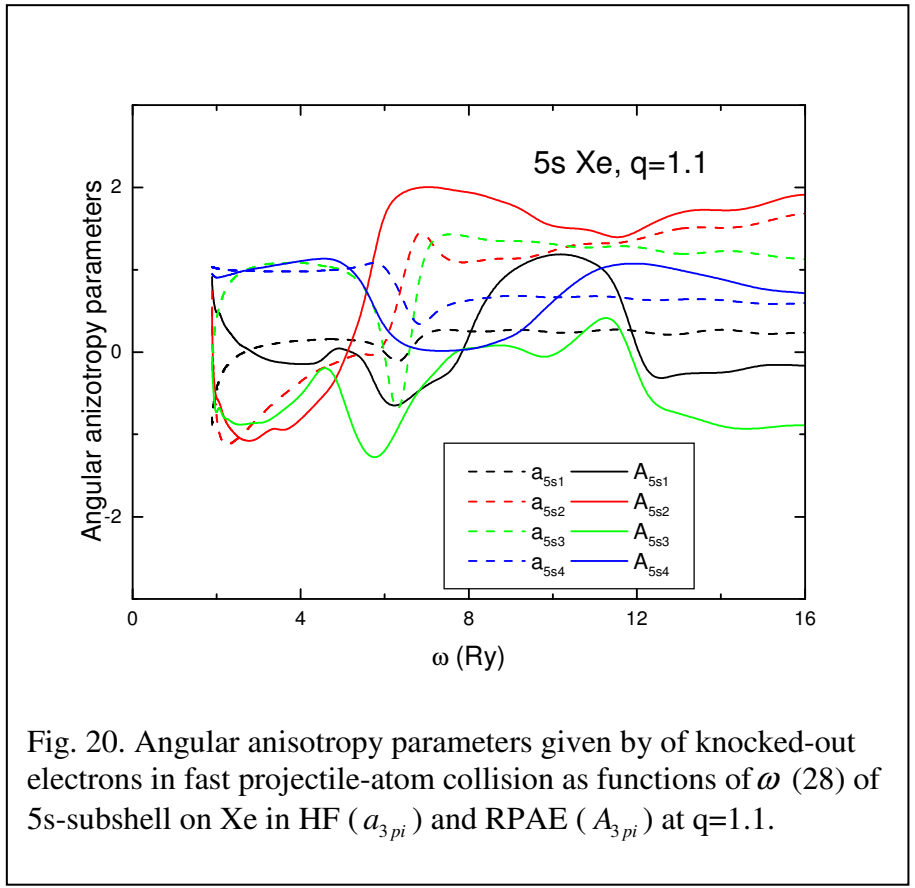

Fig. 20. Angular anisotropy parameters given by of knocked-out electrons in fast projectile-atom collision as functions of $\omega$ (28) of 5s-subshell on Xe in HF ($a_{3pi}$) and RPAE ($A_{3pi}$) at q=1.1.

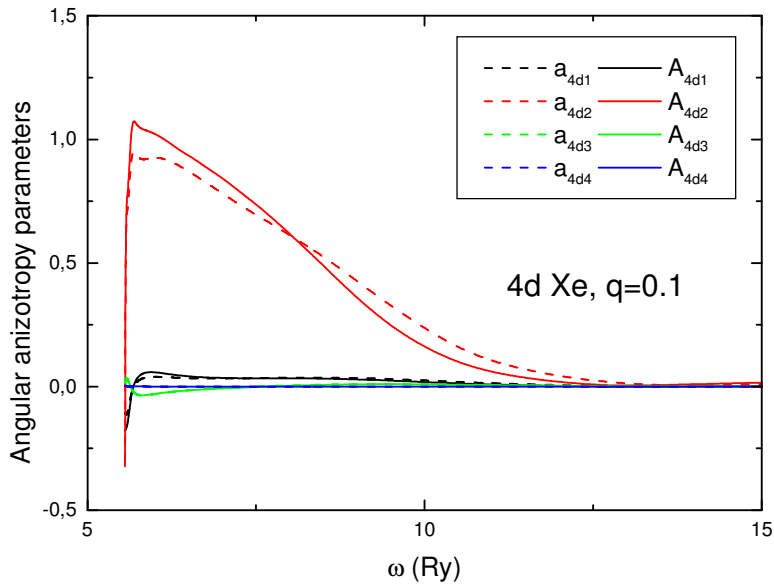

Fig. 21. Angular anisotropy parameters given by of knocked-out electrons in fast projectile-atom collision as functions of $\omega$ (28) of 4d-subshell on Xe in HF ($a_{4di}$) and RPAE ($A_{4di}$) at q=0.1.

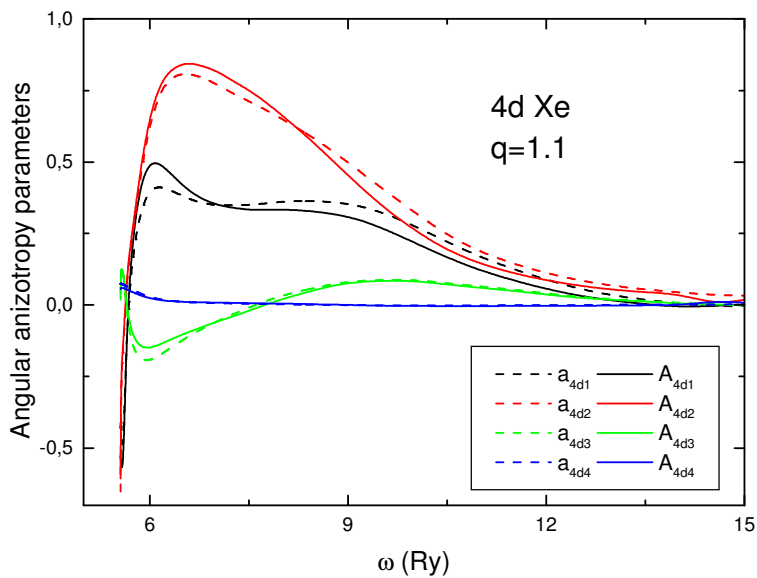

Fig. 22. Angular anisotropy parameters given by of knocked-out electrons in fast projectile-atom collision as functions of $\omega$ (28) of 4d-subshell on Xe in HF ($a_{4di}$) and RPAE ($A_{4di}$) at q=1.1.

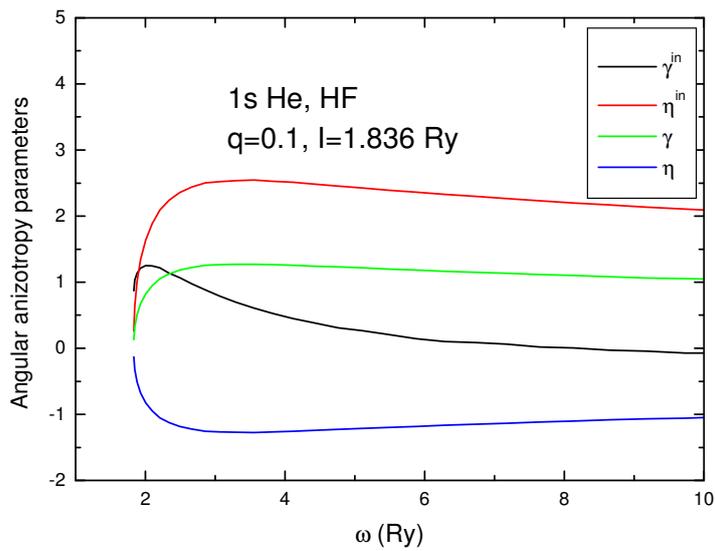

Fig. 23. Angular anisotropy non-dipole parameters of knocked-out electrons in fast projectile-atom collision in the optical limit compared to similar parameters in photoionization, given by (37, 38) at q=0.1 and (36) on He in HF

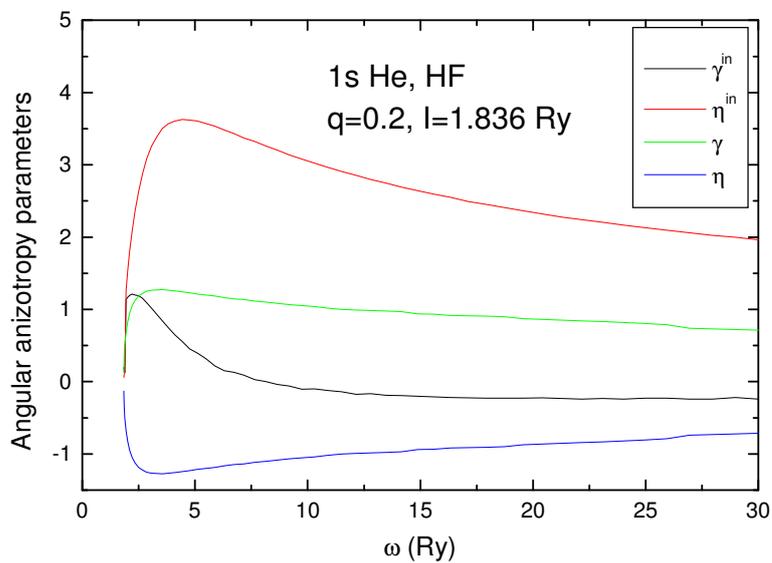

Fig. 24. Angular anisotropy non-dipole parameters of knocked-out electrons in fast projectile-atom collision in the optical limit compared to similar parameters in photoionization, given by (37, 38) at q=0.2 and (36) on He in HF.

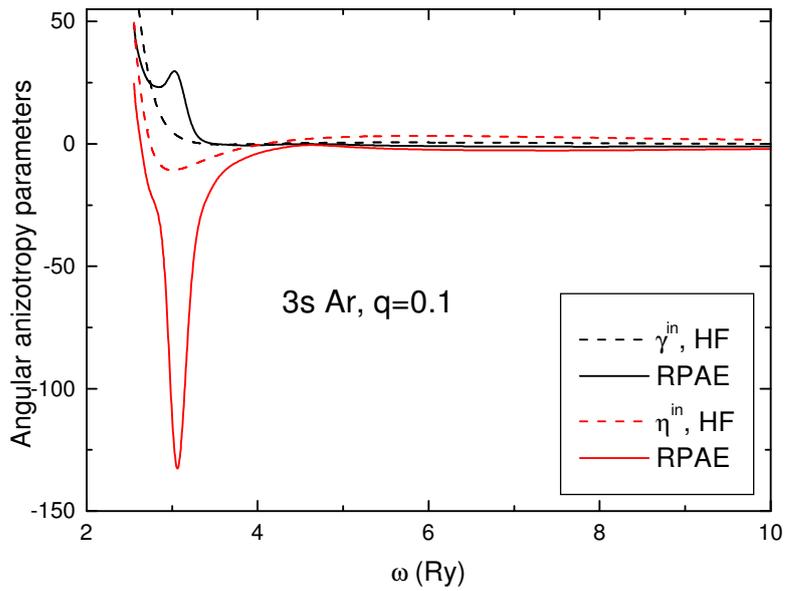

Fig. 25. Angular anisotropy non-dipole parameters of knocked-out electrons in fast projectile-atom collision in the optical limit given by (37, 38) at q=0.1 for 3s subshell of Ar.

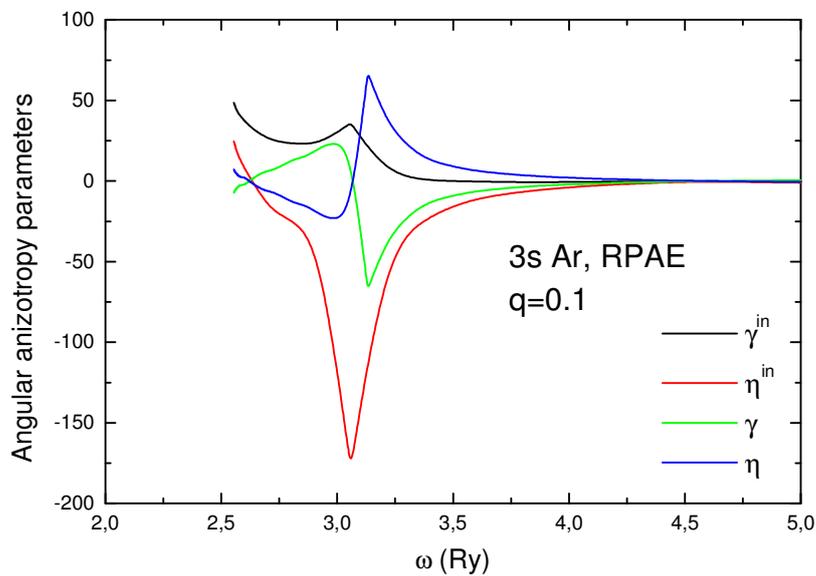

Fig. 26. Angular anisotropy non-dipole parameters of knocked-out electrons in fast projectile-atom collision in the optical limit compared to similar parameters in photoionization, given by (37, 38) at q=0.1 and (36) for 3s subshell of Ar in RPAE.

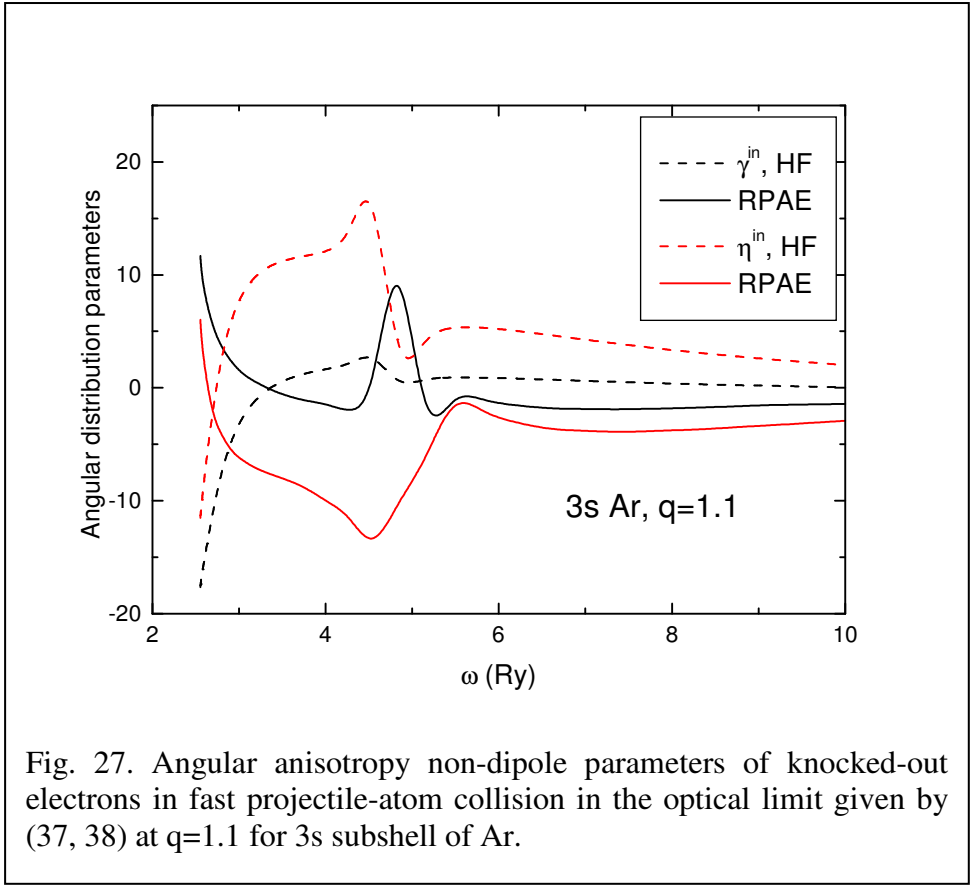

Fig. 27. Angular anisotropy non-dipole parameters of knocked-out electrons in fast projectile-atom collision in the optical limit given by (37, 38) at q=1.1 for 3s subshell of Ar.

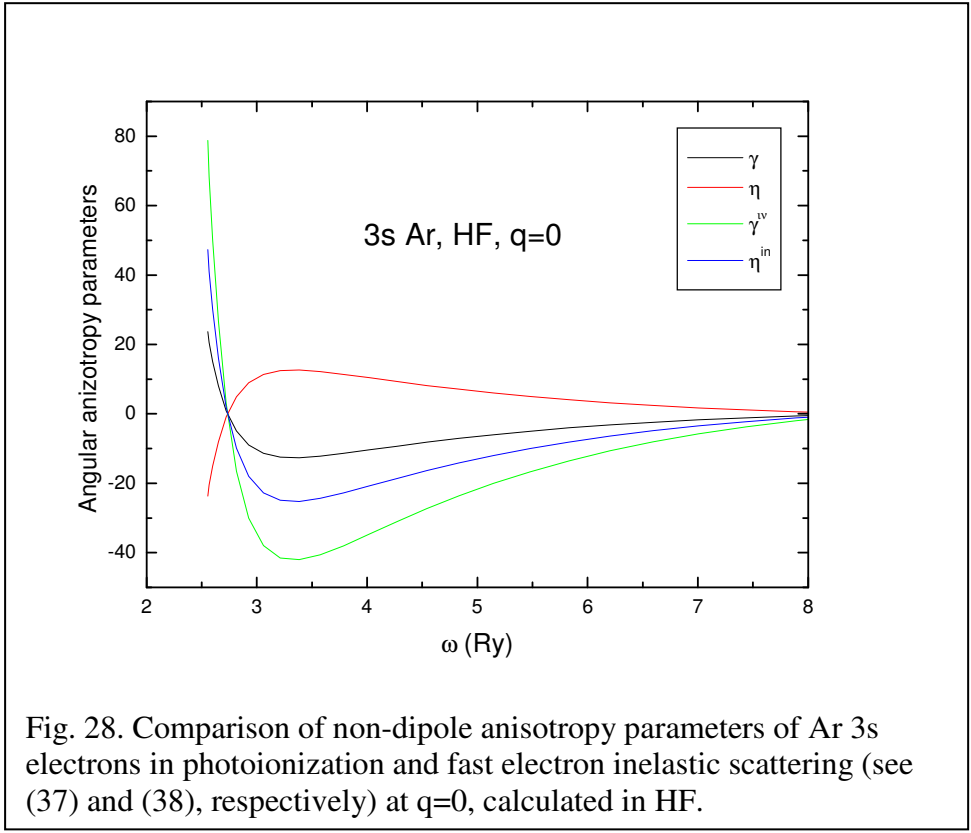

Fig. 28. Comparison of non-dipole anisotropy parameters of Ar 3s electrons in photoionization and fast electron inelastic scattering (see (37) and (38), respectively) at q=0, calculated in HF.

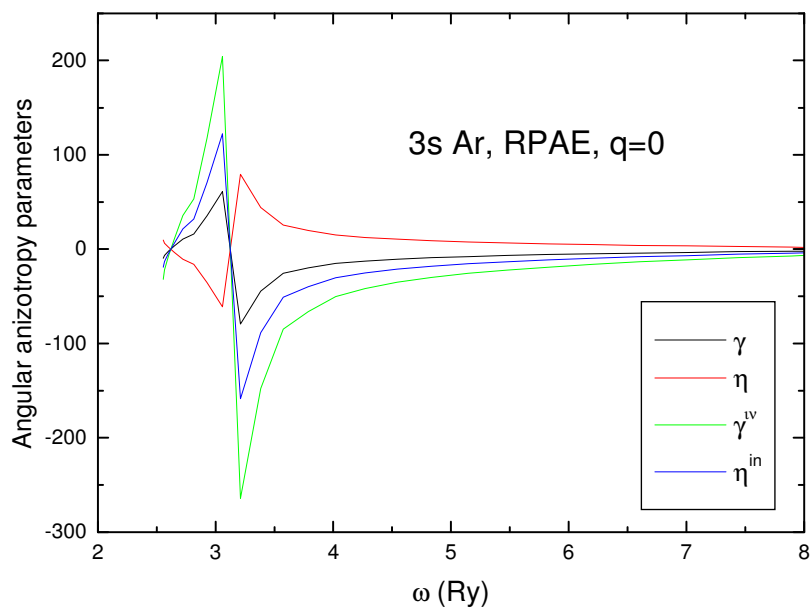

Fig. 29. Comparison of non-dipole anisotropy parameters of Ar 3s electrons in photoionization and fast electron inelastic scattering (see (37) and (38), respectively) at q=0, calculated in RPAE.

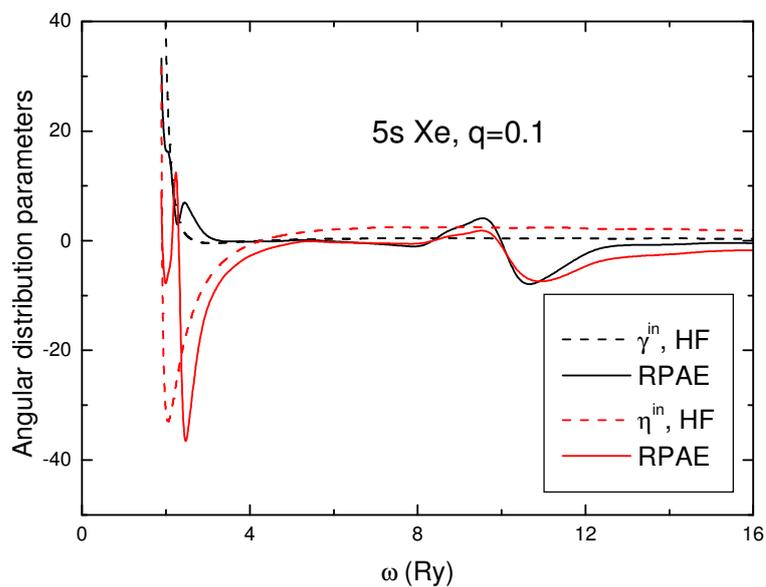

Fig. 30. Angular anisotropy parameters non-dipole parameters of knocked-out electrons in fast projectile-atom collision in the optical limit given by (37, 38) at q=0.1 for 5s subshell of Xe

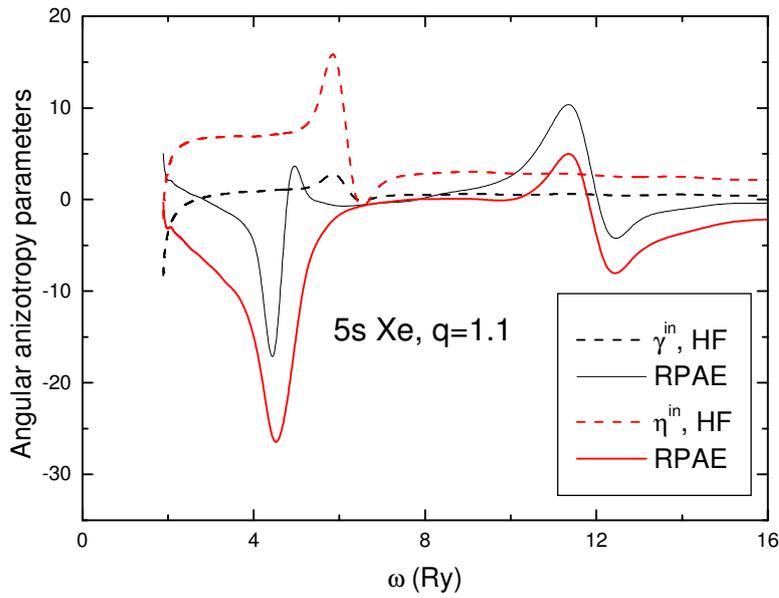

Fig. 31. Angular anisotropy non-dipole parameters of knocked-out electrons in fast projectile-atom collision in the optical limit given by (37, 38) at q=1.1 for 5s subshell of Xe

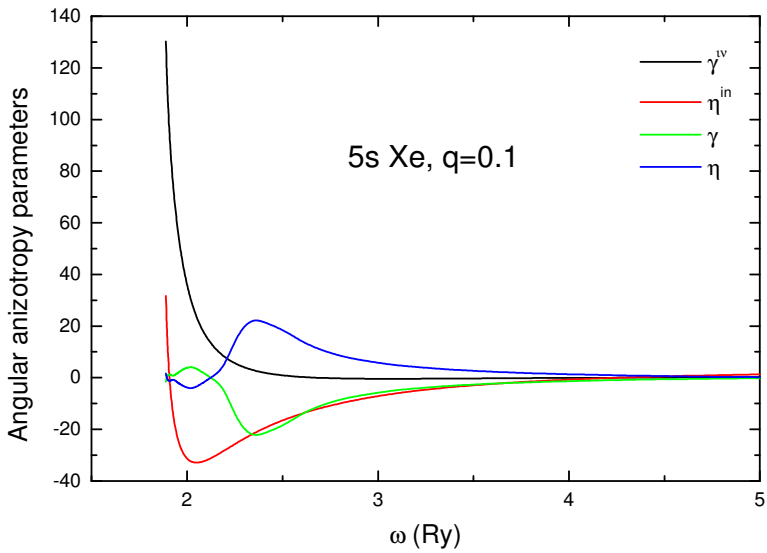

Fig. 32. Angular anisotropy non-dipole parameters of knocked-out electrons in fast projectile-atom collision in the optical limit $\gamma_{3s}^{(in)}(\omega)$ and $\eta_{3s}^{(in)}(\omega)$ given by (37, 38) at $q=0.1$ compared to similar parameters in photoionization $\gamma_{3s}(\omega)$ and $\eta_{3s}(\omega)$, given by (36) for 5s subshell of Xe in RPAE.

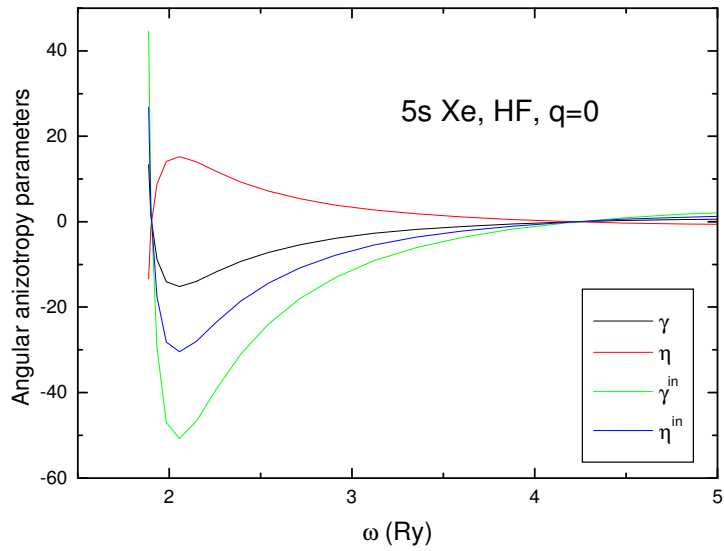

Fig. 33. Comparison of non-dipole anisotropy parameters of Xe 5s electrons in photoionization and fast electron inelastic scattering (see (37) and (38), respectively) at q=0, calculated in HF.

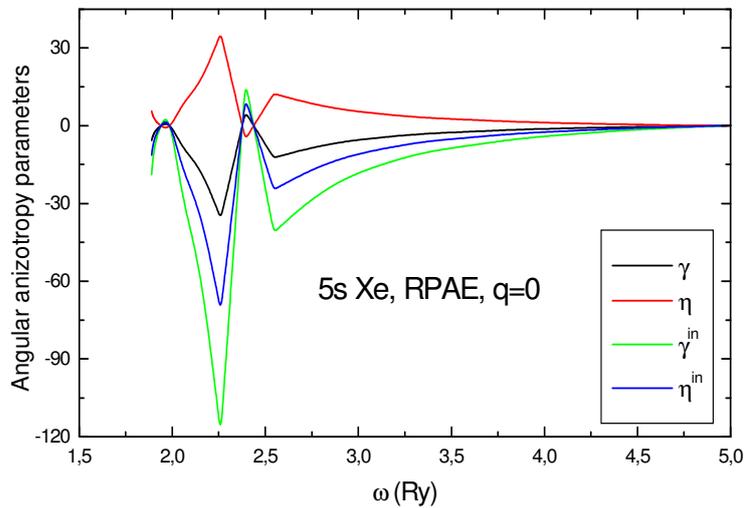

Fig 34. Comparison of non-dipole anisotropy parameters of Xe 5s electrons in photoionization and fast electron inelastic scattering (see (37) and (38), respectively) at q=0, calculated in RPAE.

## 6. Conclusions and discussion

We performed calculations for transitions from *s*, *p* and *d* subshells in three Noble gas atoms. The results demonstrate that indeed the angular anisotropy parameters are complex and informative functions, with a number of prominent variations. They depend strongly upon the outgoing electron energy and the linear momentum *q* transferred to the atom in the process of inelastic scattering of a fast electron. In Ar and Xe they are strongly affected by electron correlations.

Particular attention deserves the $q \to 0$ limit. It is seen that the non-dipole corrections to the angular distribution are essentially different for the cases of photoionization and fast electron inelastic scattering. The additional information that could come from studies of angular distribution of secondary electrons at small *q* transferred to the target in fast electron-atom collisions is of great interest and value. The suggested here experimental studies are desirable.